\input harvmac.tex
% VERSION July 25, 2003 
% VERSION August 4, 2003
% VERSION November 30, 2003
% VERSION February 10, 2004
% VERSION March, 2004
% VERSION April, 2004
%%%%%%%%%%%%%%%%%%%%%%%%%% REFERENCES %%%%%%%%%%%%%%%%%%%%%%%%%%%%%%%%%%%%%%%%%%%%%%%%%%%%

\lref\minwalla{ S. Minwalla, M. Van Raamsdonk and N. Seiberg, ``Noncommutative
Perturbative Dynamics'', JHEP {\bf 0002} (2000) 020, hep-th/9912072.}

\lref\sw{N. Seiberg and E. Witten, JHEP {\bf 9909:032} (1999).}

\lref\cds{A. Connes, M. R. Douglas, and A. Schwarz, {JHEP} {\bf 9802:003} (1998).}

\lref\douglas{M.R. Douglas and N.A. Nekrasov, Rev. Mod. Phys. {\bf 73} (2002), 977.}

\lref\szabo{ R.J. Szabo, ``Quantum Field Theory on Noncommutative Spaces'',
hep-th/0109162.}

\lref\presnajder{ P. Presnajder, ''The Origin of Chiral Anomaly and the Noncommutative
Geometry'', J. Math. Phys. {\bf 41} (2000) 2789-2804, hep-th/9912050.}

\lref\grisaru{ M.T. Grisaru and S. Penati, ''Noncommutative Supersymmetric Gauge
Anomaly'', Phys. Lett. B {\bf 504} (2001) 89-100, hep-th/0010177.}

\lref\ardalan{ F. Ardalan and N. Sadooghi, Int. J. Mod. Phys. A {\bf 16} (2001) 3151.}

\lref\gracia{J.M. Gracia-Bondia and C.P. Martin, ``Chiral Gauge Anomalies on
Noncommutative $\IR^4$'', Phys. Lett. B {\bf 479} (2000) 321.}

\lref\ardalantwo{ F. Ardalan and N. Sadooghi, ''Anomaly and Nonplanar Diagrams in
Noncommutative Gauge Theories'', Int. J. Mod. Phys. A {\bf 17} (2002) 123, hep-th/0009233.}

\lref\bonora{L. Bonora, M. Schnabl and A. Tomasiello, ``A Note on Consistent Anomalies in
Noncommutative Yang-Mills Theories'', Phys. Lett. B {\bf 485} (2000) 311, hep-th/0002210.}

\lref\bonoraone{L. Bonora and A. Sorin, ``Chiral Anomalies in Noncommutative YM Theories'',
Phys. Lett. B {\bf 521} (2001) 421, hep-th/0109204.}

\lref\langmann{E. Langmann and J. Mickelsson, ``Anomalies and Schwinger Terms in
Noncommutative Gauge Field Theory Models'', J. Math. Phys. {\bf 42} (2001) 4779,
hep-th/0103006.}

\lref\mink{C.P. Martin, ``Chiral Gauge Anomalies on Noncommutative Minkowski Spacetime'', Mod.
Phys. Lett. A {\bf 16} (2001) 311.}

\lref\intri{K. Intriligator and J. Kumar, ``$\star$-Wars Episode I: The Phantom Anomaly'',
hep-th/0107199.}

\lref\naka{T. Nakajima, ``Conformal Anomalies in Noncommutative Gauge Theories'', hep-th/0108158.}

\lref\nakados{T. Nakajima, ``UV/IR Mixing and Anomalies in Noncommutative Gauge Theories'',
hep-th/0205058.}

\lref\guts{P. Aschieri, B. Jurco, P. Schupp and J. Wess, ``Noncommutative GUT's Standard
Model and CPT'', hep-th/0205214.} 

\lref\cpmartin{ C.P. Martin, ''The Covariant Form of the Gauge Anomaly on Noncommutative
$\IR^{2N}$'', Nucl. Phys. B {\bf 623} (2002) 150-164, hep-th/0110046.}

\lref\banerjee{ R. Banerjee and S. Ghosh, ''Seiberg-Witten Map and the Axial
Anomaly in Noncommutative Field Theory'', Phys. Lett. B {\bf 533} (2002) 162-167,
hep-th/0110177.}

\lref\sadooghi{N. Sadooghi and M. Mohammadi, ''On the Beta Function and Conformal
Anomaly of Noncommutative QED with Adjoint Matter Fields'', hep-th/0206137.}

\lref\cpmartinsw{ C.P. Martin, ''The Gauge Anomaly and the Seiberg-Witten Map'',
hep-th/0211164.}

\lref\rrcouplings{ R. Banerjee, ``Anomalies in Noncommutative Gauge Theories,
Seiberg-Witten Transformation and Ramond-Ramond Couplings'',  Int. J. Mod. Phys. A {\bf 19}
(2004) 613, hep-th/0301174.}

\lref\calmet{X. Calmet and M. Wohlgenannt, ``Effective Field Theories on Noncommutative
Space-time'',  Phys. Rev. D {\bf 68} (2003) 025016, hep-ph/0305027.}

\lref\brandt{F. Brandt, C.P. Martin and F. Ruiz Ruiz, ``Anomaly Freedom in Seiberg-Witten
Noncommutative Gauge Theories'', JHEP (2003) {\bf 0307:068}, hep-th/0307292.}

\lref\review{F. Brandt, ``Seiberg-Witten Maps and Anomalies in Noncommutative Yang-Mills
Theories'', hep-th/0403143.}

\lref\cham{A.H. Chamseddine, Commun. Math. Phys. {\bf 218} (2001) 283.}

\lref\chamone{A.H. Chamseddine, {Phys. Lett.} B {\bf 504} (2001) 33.}

\lref\chamtwo{A.H. Chamseddine, ``Invariant Actions for Noncommutative
Gra\-vi\-ty'', hep-th/02\-02\-137.}

\lref\moffat{J.W. Moffat, {Phys. Lett.} B {\bf 491} (2000) 345; {Phys.
Lett.} B {\bf 493} (2000) 142.}

\lref\moffatperturb{J.W. Moffat, ``Perturbative Noncommutative Quantum Gravity'',
hep-th/0008089.}

\lref\daniela{M.A. Cardella and D. Zanon, ``Noncommutative Deformation of
Four-dimensional Einstein Gravity'', hep-th/0212071.}

\lref\maxim{  M. Kontsevich, ``Deformation Quantization of Poisson
Manifolds I'', q-alg/9709040.}

\lref\nctopo{H. Garcia-Compe\'an, O. Obreg\'{o}n, C. Ramirez, and
M. Sabido, ``Noncommutative Topological Theories of Gravity'', Phys. Rev. D {\bf 68} (2003) 
045010, hep-th/0210203.}

\lref\ncsdg{H. Garcia-Compe\'an, O. Obreg\'{o}n, C. Ramirez, and
M. Sabido, ``Noncommutative Self-dual Gravity'', Phys. Rev. D {\bf 68} (2003) 044015,
hep-th/0302180.}

\lref\salam{ R. Delbourgo and A. Salam, ``The Gravitational Correction to PCAC'', Phys. Lett.
{\bf 40B} (1972) 381.}

\lref\freund{T. Eguchi and P.G.O. Freund, ``Quantum Gravity and World Topology'',  Phys. Rev.
Lett. {\bf 37} (1976) 1251.}

\lref\delbourgo{ R. Delbourgo, ``A Dimensional Derivation of the Gravitational PCAC
Correction'', J. Phys. A: Math. Gen. {\bf 10} (1977) L237.}

\lref\alwitt{L. Alvarez-Gaum\'e and E. Witten, Nucl. Phys. B {\bf 234} (1983) 269.}

\lref\ginsparg{L. Alvarez-Gaum\'e and P. Ginsparg, ``The Topological Meaning of Nonabelian
Anomalies'', Nuc. Phys. B {\bf 243} (1984) 449; The Structure og Gauge and Gravitational
Anomalies'', Ann. Phys. {\bf 161} (1985) 423.}

\lref\globo{E. Witten, Commun. Math. Phys. {\bf 100} (1985) 197.}

\lref\harvey{ J.A. Harvey, ``Topology of the Gauge Group in Noncommutative Gauge Theory'',
hep-th/0105242.}

\lref\kuiper{ N.H. Kuiper, Topology {\bf 3} (1965) 19; R.S. Palais, Topology {\bf 3} (1965)
271.}

\lref\schw{J. Schwinger, Phys. Rev. D {\bf 82} (1951) 664.}

\lref\madoregrosse{H. Grosse and J. Madore, ``A Noncommutative Version of the Schwinger
Model'', Phys. Lett. B {\bf 283} (1992) 218.}

\lref\grosse{H. Grosse and P. Presnajder, ``A Noncommutative Regularization of the
Schwinger Model'',  Lett. Math. Phys. {\bf 46} (1998) 61; ``A Treatment of the Schwinger
Model within Noncommutative Geometry'', hep-th/9805085.}

\lref\bala{A.P. Balachandran and  S. Vaidya, ``Instantons and Chiral Anomaly in Fuzzy
Physics'', Int. J. Mod. Phys. A {\bf 16} (2001) 17, hep-th/9910129.}

\lref\ydri{B. Ydri, ``Noncommutative Chiral Anomaly and the Dirac-Ginsparg-Wilson
Operator'', JHEP {\bf 0308} (2003) 046, hep-th/0211209.}

\lref\wess{B. Jurco, S. Schraml, P. Schupp and J. Wess, {Eur. Phys. J. C} {\bf 17}
(2000) 521; B. Jurco, P. Schupp and J. Wess, {Nucl. Phys.} B {\bf 604}
(2001) 148; J. Wess, {Commun. Math. Phys.} {\bf 219} (2001) 247; B. Jurco, L. Moller, S. Schraml,
P. Schupp and J. Wess, {Eur. Phys. J. C} {\bf 21} (2001) 383; X. Calmet, B. Jurco, P. Schupp, J.
Wess and M. Wohlgenannt, {\it Eur. Phys. J. C} {\bf 23} (2002) 363.}

\lref\sheikh{A. Micu and M.M. Sheikh-Jabbari, ``Noncommutative $Phi^4$ Theory at Two Loops'',
JHEP {\bf 0101} (2001) 025, hep-th/0008057.}

\lref\esperanza{A. Armoni, E. Lopez and S. Theisen, ``Nonplanar Anomalies in Noncommutative
Theories and the Green-Schwarz Mechanism'', JHEP {\bf 0206:050} (2002), hep-th/0203165.}

\lref\calmet{X. Calmet and M. Wohlgenannt, ``Effective Field Theories on Non-Commutative
Space-Time'', Phys. Rev. D {\bf 68} (2003) 025016, hep-ph/0305027.}

\lref\perrot{ D. Perrot, ``In the Topological Interpretation of Gravitational Anomalies'', J.
Geom. Phys. {\bf 39} (2001) 82, math-ph/0006003.}

%%%%%%%%%%%%%%%%%%%%%%%%%% FIGURES%%%%%%%%%%%%%%%%%%%%%%%%%%%%%%%

\let\includefigures=\iftrue
\newfam\black
\includefigures
\input epsf
\def\figin{\epsfcheck\figin}\def\figins{\epsfcheck\figins}
\def\epsfcheck{\ifx\epsfbox\UnDeFiNeD
\message{(NO epsf.tex, FIGURES WILL BE IGNORED)}
\gdef\figin##1{\vskip2in}\gdef\figins##1{\hskip.5in}% blank space instead
\else\message{(FIGURES WILL BE INCLUDED)}%
\gdef\figin##1{##1}\gdef\figins##1{##1}\fi}
\def\DefWarn#1{}
\def\figinsert{\goodbreak\midinsert}
\def\ifig#1#2#3{\DefWarn#1\xdef#1{fig.~\the\figno}
\writedef{#1\leftbracket fig.\noexpand~\the\figno}%
\figinsert\figin{\centerline{#3}}\medskip\centerline{\vbox{\baselineskip12pt
\advance\hsize by -1truein\noindent\footnotefont{\bf Fig.~\the\figno:} #2}}
\bigskip\endinsert\global\advance\figno by1}
%%%
\else
\def\ifig#1#2#3{\xdef#1{fig.~\the\figno}
\writedef{#1\leftbracket fig.\noexpand~\the\figno}%
%\figinsert\figin{\centerline{#3}}\medskip\centerline{\vbox{\baselineskip12pt
%\advance\hsize by -1truein\noindent\footnotefont{\bf Fig.~\the\figno:} #2}}
%\bigskip\endinsert
\global\advance\figno by1}
\fi

%%%%%%%%%%%%%%%%%%%%%  Math-style letters   %%%%%%%%%%%%%%%%%%%%%%%%
\font\cmss=cmss10 \font\cmsss=cmss10 at 7pt

\def\IB{\relax\hbox{$\inbar\kern-.3em{\rm B}$}}
\def\IC{\relax\hbox{$\inbar\kern-.3em{\rm C}$}}
\def\IQ{\relax\hbox{$\inbar\kern-.3em{\rm Q}$}}
\def\ID{\relax\hbox{$\inbar\kern-.3em{\rm D}$}}
\def\IE{\relax\hbox{$\inbar\kern-.3em{\rm E}$}}
\def\IF{\relax\hbox{$\inbar\kern-.3em{\rm F}$}}
\def\IG{\relax\hbox{$\inbar\kern-.3em{\rm G}$}}
\def\IGa{\relax\hbox{${\rm I}\kern-.18em\Gamma$}}
\def\IH{\relax{\rm I\kern-.18em H}}
\def\IK{\relax{\rm I\kern-.18em K}}
\def\IL{\relax{\rm I\kern-.18em L}}
\def\IP{\relax{\rm I\kern-.18em P}}
\def\IR{\relax{\rm I\kern-.18em R}}
\def\Z{\relax\ifmmode\mathchoice
{\hbox{\cmss Z\kern-.4em Z}}{\hbox{\cmss Z\kern-.4em Z}}
{\lower.9pt\hbox{\cmsss Z\kern-.4em Z}}
{\lower1.2pt\hbox{\cmsss Z\kern-.4em Z}}\else{\cmss Z\kern-.4em
Z}\fi}
\def\IZ{Z\!\!\!Z}
\def\II{\relax{\rm I\kern-.18em I}}

%%%%%%%%%%%%%%%%%%%%% Calligraphic letters  %%%%%%%%%%%%%%%%%%%%%

%%%%%%%%%%%%%%%%%%%%%%%%%% Derivatives  %%%%%%%%%%%%%%%%%%%%%%%%

%%%%%%%%%%%%%%%%%%%% letters with bar %%%%%%%%%%%%%%%%%%%%%%%%%%

\def\hat{\widehat}
\def\bar{\overline}

%%%%%%%%%%%%%%%%%%%%%%%%%%% Math symbols %%%%%%%%%%%%%%%%%%%%%%%

%\def\np{\nabla_{\partial}}
%\def\npb{\nabla_{\bar {\partial}}}

%\def\clieg{{\bf g}_{\scriptscriptstyle{\IC}}}
%\def\cliet{{\bf t}_{\scriptstyle{\IC}}}
%\def\cliek{{\bf k}_{\scriptscriptstyle{\IC}}}
%\def\clies{{\bf s}_{\scriptstyle{\IC}}}

\def\inbar{\,\vrule height1.5ex width.4pt depth0pt}

%%%%%%%%%%%%%%%%%%%   Greek letters %%%%%%%%%%%%%%%%%%%

\def\bar{\overline}

\def\example#1{\bgroup\narrower\footnotefont\baselineskip\footskip\bigbreak
\hrule\medskip\nobreak\noindent {\bf Example}. {\it #1\/}\par\nobreak}
\def\endexample{\medskip\nobreak\hrule\bigbreak\egroup}

%%%%%%%%%%%%%%%%%%%%%%%%%%%%%%%%%%%%%%%%%%%%%%%%%%%%%%%%%%%%%%%%%%%%%%%%%%%%%%%
%%%%%%%%%%%%%%%%%%% TITLE PAGE  %%%%%%%%%%%%%%%%%%%%%%%%%%%%%%%%

\Title{\vbox{\baselineskip12pt
\hbox{CINVESTAV-FIS-13/04}
\hbox{hep-th/yymmddd}
}} {\vbox{
\centerline{Gravitational Anomalies in}
\centerline{Noncommutative Field Theory}}}
\centerline{Sendic Estrada-Jim\'enez\foot{E-mail address: sendic@fis.cinvestav.mx}, Hugo
Garcia-Compean\foot{{\it ASICTP Associate
Member, The Abdus Salam 
International Centre for Theoretical Physics, Trieste, Italy}, E-mail address:  
compean@fis.cinvestav.mx} and  Carlos Soto-Campos\foot{E-mail address: csoto@fis.cinvestav.mx}}
%\medskip
%\medskip
%\medskip
\medskip
%\vskip 8pt
\centerline{\it Departamento de Fisica}
\centerline{\it Centro de Investigaci\'on y de Estudios Avanzados del IPN}
\centerline{\it Apdo. Postal 14-740, 07000, M\'exico D.F., M\'exico}
%\medskip
%\medskip
\bigskip

\vskip 10pt

\vskip 20 pt

{\bf \centerline{Abstract}}
\noindent
Gravitational axial and chiral anomalies in a noncommutative space are examined through the
explicit perturbative computation of one-loop diagrams in various dimensions. The analysis
depend on how gravity is coupled to noncommutative matter fields. Delbourgo-Salam computation
of the gravitational axial anomaly contribution to the pion decay into two photons, is studied
in detail in this context. In the process we show that the two-dimensional chiral pure
gravitational Weyl anomaly does not receive noncommutative corrections. Pure gravitational
chiral anomaly in $4k+2$ dimensions with matter fields being chiral fermions of spin-${1 \over
2}$ and spin-${3 \over 2}$, is discussed and a noncommutative correction is found in both
cases. Mixed anomalies are finally considered for both cases.

\smallskip

%PACS numbers: {\it 03.50.-z, 03.50.De, 11.10.-z, 03.65.Db}

%\medskip
\Date{April, 2004}

%%%%%%%%%%%%%%%%%%%%%%%%%%%%%%%%%%%%%%%%%%%%%%%%%%%%%%%%%%%%%%%%%%%%
%%%%%%%%%%%%%%%%%%%%%%%%%%%%%%%%%%%%%%%%%%%%%%%%%%%%%%%%%%%%%%%%%%%%%

\newsec{Introduction}

Noncommutative field theory has very intriguing new effects in quantum field theory as the
UV/IR mixing discovered recently in Ref. \minwalla, which in fact, has a stringy origin. Other
nice surprise is the deep relation with string theory and M-theory \refs{\sw,\cds} (for a nice
review see, for instance, \refs{\douglas,\szabo}). Another important effect in quantum field
theory are the anomalies. Gauge anomalies, in particular axial and gauge chiral anomalies in
noncommutative gauge theories has been discussed in a series of papers by various authors
\refs{\presnajder,\grisaru,\ardalan,\gracia,\ardalantwo,\bonora,\bonoraone,\langmann,\mink,\intri,
\naka,\nakados,\guts,\cpmartin,\banerjee,\sadooghi,\cpmartinsw, \rrcouplings,\brandt,\review}.
In particular, noncommutative gauge anomalies in noncommutative Yang-Mills theory was
considered in Refs. \refs{\gracia,\bonora,\bonoraone,\mink,\intri} for planar diagrams with
gauge group $U(N)$.  For the case of non-planar diagrams there has been some previous work in
\refs{\ardalantwo,\esperanza,\nakajima}. The analysis can be extended to other gauge groups by
introducing the Seiberg-Witten map as in references
\refs{\banerjee,\cpmartinsw,\rrcouplings,\brandt,\review}.

%%%%%%%%%%%%%%%%%%%%%%%%%%%%%%%%%%%%%%%%%%%%%%%

On the other hand, recently various noncommutative theories of gravity have been proposed. In
particular, the proposals in Refs. \refs{\cham,\chamone,\chamtwo,\moffat,\moffatperturb,
\daniela}, different Moyal deformations of Einstein gravity in four dimensions are
given. All these actions however are not manifestly invariant under the full noncommutative
transformations since they are deformed under the Moyal product, with a constant
noncommutativity parameter. Therefore they are not diffeomorphism invariant as far the Moyal
product
depend on the coordinate system. These products can be made diffeomorphism invariant,
substituting the Moyal $*_M$-product by the Kontsevich $*_K$-product \maxim.  In this paper
we will
assume the uses of the Kontsevich $*_K$-product though we will avoid to use the
subscript $K$.  

Recently, another noncommutative proposals are given in Ref. \refs{\nctopo,\ncsdg}. In the
former reference a manifestly SO(1,3)
invariant noncommutative topological action for the gravitational theta terms is
constructed. For appropriate boundary conditions give us the possibility to provide some
insight about noncommutative gravitational instantons and noncommutative Lorentz
gravitational anomalies.
In the
latter paper it was discussed a dynamical case
of Einstein gravity by Moyal (or Kontsevich) deforming the self-dual projection of Einstein
theory to find a manifestly SL$(2,\IC)$ invariant noncommutative theory. 

Noncommutative topological actions which are the linear combination of the Euler number
$\widehat{\chi}(X)$ and signature $\widehat{\sigma}(X)$, being an $\widehat{{\rm
SO}(3,1)}$-invariant action, are very important since they describe the breakdown of chiral
symmetry in the presence of gravitational fields. Due the technical difficulties in trying
to make noncommutative the action $\chi(X)$ we propose a way by getting the noncommutative
$\widehat{{\rm SL}(2,\IC)}$-invariant action $\widehat{\chi}(X)$, from a noncommutative
version for the signature $\widehat{\sigma}(X)$, which is also, a $\widehat{{\rm
SL}(2,\IC)}$-invariant action.

On the other hand, local gauge anomalies are associated to the lack of invariance of the
fermionic
one-loop effective action $\Gamma(Q) = \log \big[\det \not \! \! D\big]$, with $e^{-\Gamma} =
\int
{\cal D} \psi {\cal D} \psi^* e^{- \int_X L}$, under infinitesimal gauge transformations
with the chiral matter fields $\psi(x)$ and $\psi^*(x)$ living in a complex
representation $Q$ of the gauge group $G$. In the case of theories gravitational couplings to
matter there are different types of gravitational anomalies, depending on the type of
transformations. Thus the Lorentz (or
automorphisms) anomaly is related to the lack of gauge invariance of $\Gamma$ under Lorentz 
transformations. When the symmetry group are the group of diffeomorphisms $Diff(X)$ of a
smooth spacetime manifold $X$ and $\Gamma(Q)$ is not generally covariant, then we have a
diffeomorphism anomaly. 

Gravitational corrections to the ABJ-anomaly was originally computed by Delbourgo and
Salam \refs{\salam} and further worked out in Refs. \refs{\freund,\delbourgo}. Soon after, 
gravitational anomalies were computed in a systematic way by Alvarez-Gaum\'e and
Witten \alwitt\ (see also \ginsparg). Through out the present paper we will not consider
global gravitational
anomalies \globo.

In Ref. \nctopo\ it was argued about a noncommutative version of the the Lorentz group
$\widehat{SO(4)}$ following a global procedure for computing chiral anomalies in gauge
theory suggested by Harvey \harvey, which is based in the mathematical literature \kuiper.
The application of
these ideas to the diffeomorphism transformations connected to the identity, might predict
new nontrivial noncommutative gravitational effects, which should be computed explicitly
as a noncommutative correction to the gravitational contribution to the chiral anomaly.

In the present paper we compute gravitational axial and chiral anomalies starting from a of
the full noncommutative gravitational theory and focus in the interaction lagrangian between
chiral fermions and gravitons in the noncommutative spacetime. We will follow the t' Hooft's
observation that the anomalies can be understood in terms of the low energy effective field
theory and we will consider a noncommutative effective field theory describing the one-loop
effective action of chiral fermions in background fields being the noncommutative
gravitational field (pure gravitational anomalies) or/and noncommutative Yang-Mills fields
(mixed anomalies). we will restrict to the computation of perturbative one-loop diagrams of
chiral fermions with external gravitons or/and gluons in various dimensions. Only planar
diagrams are considered in this paper.

This paper is organized as follows: In Section 2 we start with some arguments about some
global aspects on noncommutative gravitational anomalies. In Section 3 we provide some basic
features of perturbative noncommutative gravity and the corresponding noncommutative Feynman
rules of the coupling of Weyl fermions to gravity. Section 4 is devoted to compute the
noncommutative analog of Delbourgo-Salam axial gravitational anomaly which is the
gravitational correction to the ABJ (axial) anomaly in four dimensions.

In Section 5 we discuss the gravitational gauge chiral anomaly in two dimensions. We show that
in this
case the noncommutative case coincides with the commutative one and there is not
noncommutative correction. Section 6 is devoted to
extend the
computation of the one-loop amplitude to dimension $D=4k+2$. In here, after some preliminaries
we compute the gravitational chiral gauge anomaly by evaluating directly the
perturbative amplitude of the one loop
diagram and the Schwinger procedure. In the same section is also computed the amplitude for
spin-${3 \over 2}$ chiral fermions. In Section 7 we
describe separately the mixed
anomalies between gauge fields and gravitational fields coupled with spin-${1
\over 2}$ or spin-${3 \over 2}$ in a noncommutative space.
Finally in Section 8 we give our final remarks. 

\vskip 1truecm
%%%%%%%%%%%%%%%%%%%%%%
\noindent

\newsec{Towards Noncommutative Gravitational Anomalies: Global Aspects}

Before to proceed to compute gravitational anomalies in the noncommutative context we would
like to make some global considerations about the nature of these anomalies.

From the topological perspective local gravitational anomalies are obtained through the
computation of some suitable homotopy groups of the relevant gauge group. 

In Ref. \nctopo\ it was argued about a noncommutative version of the the Lorentz group
$\widehat{SO(4)}$ following a global procedure for computing chiral anomalies in gauge
theory suggested by Harvey \harvey, which is based in the mathematical literature \kuiper.
The proposal consists in assuming that $\widehat{SO(4)}$
consist of the set of compact orthogonal operator algebra ${\bf O}_{cpt}({\cal H})$, defined
on the separable real Hilbert space ${\cal H}$.  The compactness property avoids the Kuiper
theorem, which states that the set of pure orthogonal operators ${\bf O}({\cal H})$ has
trivial homotopy groups \kuiper. This algebra has non-trivial subalgebras which have the same
homotopy than $SO(\infty)$ (up to Bott's periodicity 8), which may give rise to nontrivial new
topological effects in noncommutative gravity. Noncommutative local Lorentz anomaly is
detected with $\pi_{3}({\bf O}_{cpt}({\cal H})) = \IZ$. The choice of ${\bf O}_{cpt}({\cal
H})$ as a version of $\widehat{SO(4)}=SO(\infty)$ is highly not unique, thus there are many
possibilities to do that and there is not a natural procedure to define $\widehat{SO(4)}$
and a more explicit way of computing the local gravitational anomaly is indeed needed.

In gravitational theories, the Lorentz group is only a part of the entire symmetry group.
Thus, the moduli space of the pure gravity theory
involves a richer phase space structure which consist of the quotient space: ${\cal M} = {\cal
T}/\Gamma^+_{\infty}$, where
${\cal T}= Met(X)/Diff^+_0(X)$ is the Teichm\"uller space and $\Gamma^+_{\infty}$ is the
mapping
class group given by the quotient group $\Gamma^+_{\infty} = Diff^+(X)/Diff^+_0(X)$. Here $Met(X)$
is the moduli
space of Riemannian metrics on $X$, $Diff^+(X)$ is the group of all
orientation preserving diffeomorphisms on $X$ and
$Diff^+_0(X)$ is the group of orientation preserving diffeomorphisms on $X$ which
are connected to the identity. However there is a restriction in the spacetime dimensionality in
which the diffeomorphism  anomaly can exists. This can exist only for $dim X =4k + 2$
dimensions since only in $D=4k+2$ dimensions, the orthogonal group $O(1,D-1)$ has complex
representations.

Local gravitational anomalies in the usual commutative case appear when the mapping class group is
the trivial group i.e., $\Gamma^+_{\infty}= 1$. Thus the moduli space is given by ${\cal M} =
Met(X)/Diff^+_0(X)$. The global gravitational anomalies are related to the
disconnectedness of
$\Gamma^+_{\infty}$, i.e. $\pi_0(\Gamma^+_{\infty}) \not= 1$. Now the moduli space for
noncommutative gravity might be defined by  $\widehat{\cal M} =
\widehat{\cal T}/ \widehat{\Gamma^+_{\infty}}$ with $\widehat{{\cal T}}=
\widehat{Met(X)}/\widehat{Diff^+_0(X)}$ and $\widehat{\Gamma^+_{\infty}} =
\widehat{Diff^+(X)}/\widehat{Diff^+_0(X)}$. Of course, on order to perform some computations
on anomalies, one has to be able to provide
suitable definitions for $\widehat{\Gamma^+_{\infty}},$ $\widehat{Diff^+(X)}$ and
$\widehat{Diff^+_0(X)}$. Noncommutative local gravitational anomalies would arise when 
$\pi_2(\widehat{\cal M}) = \pi_1(\widehat{Diff^+_0(X)})\not=1$, where 
$\widehat{\cal M} =
\widehat{Met(X)}/\widehat{Diff^+_0(X)}$.

Once again the choice of some suitable version of $\widehat{Diff^+_0(X)}$ 
as is highly not unique, there are many
possibilities for it and there is not a natural procedure to define
$\widehat{Diff^+_0(X)}$
and a more explicit way of computing the local gravitational anomaly is needed. In this
paper we avoid the use the topological perspective and we will compute gravitational chiral
anomalies by the direct and explicit computation of one-loop diagrams of chiral fermions coupled to
external
gravitons and/or gauge fields. In order to do that we will use the
Feynman
rules of a suitable noncommutative gravity given in the next section. 

\vskip 1truecm
%%%%%%%%%%%%%%%%%%%%%%
\noindent

\newsec{The Noncommutative Coupling of Gravity and Chiral Fermions}

In this section we will give a brief overview of the pure noncommutative
perturbative gravitational field and
interaction with a noncommutative Weyl fermion. Our aim is to recall the relevant structure of
couplings and Feynman rules, which we will need in the following sections.

As we reviewed in the introduction, at present there is not well established a definitive 
and well defined realistic noncommutative gravity theory. In this paper we will not deal 
with any specific noncommutative theory of gravity. This is because at the end we will not
consider an specific theory of pure gravity but we will be interested only in the interactions of
linearized noncommutative gravitational field to
chiral fermions. However to be concrete we will briefly review a particular proposal 
of noncommutative Einstein gravity \chamone\ given by the noncommutative Einstein-Hilbert
action: $\widehat{I}_{EH}=   
-{1\over 16 \pi G_N} \int_X d^4x (-e)* e_{\mu}^a(x) * e_{\nu}^b(x) * R^{\mu \nu}_{ab}(x),$
where $g_{\mu \nu}(x) = e_{\mu}^a(x) * e_{\nu}^b(x) \eta_{ab}$, and 
$R^{ab}_{\mu \nu}(x) = \partial_{\mu} \omega_{\nu}^{ab}(x) - 
\partial_{\nu} \omega_{\mu}^{ab}(x) + [\omega_{\mu}(x), \omega_{\nu}(x)]_*^{ab}$, with 
$\omega_{\mu}^{ab}(x)$, being the noncommutative spin connection and $[A,B]_*
\equiv
A*B - B*A$ is the Moyal bracket.
Here $*$-product is defined by $F*G(x)\equiv  \exp \bigg({i \over 2}\Theta^{\mu \nu}{\partial
\over \partial y^{\mu}}{\partial  \over \partial z^{\nu}} \bigg) 
F(y)G(z)\bigg|_{y=z=x}.$ From now on in order to avoid causality problems we will
take $\theta^{0\nu} = 0.$

Noncommutative perturbative gravity is defined as by a perturbative expansion $I= I^{(0)} +
I^{(1)} + I^{(2)} +  {\cal O}(\kappa^4)$ of the noncommutative Einstein-Hilbert action 
\moffatperturb\ generated by a perturbative  expansion
of the metric as follows $g_{\mu \nu} = \eta_{\mu \nu} - \kappa h_{\mu \nu} + \kappa^2
h_{\mu}^{\alpha}* h_{\alpha \nu}
- \kappa^3 h_{\mu}^{\alpha} * h_{\alpha \beta} * h_{\nu}^{\beta} + {\cal O}(\kappa^4).$

In Ref. \moffatperturb\ it was given the Feynman rules of this pure noncommutative gravity theory.
In what follows we will give the corresponding Feynman rules governing the coupling of the
noncommutative linear metric $h_{\mu \nu}(x)$ to chiral fermions.

%%%%%%%%%%%%%%%%%%%%%%%%%%%%%%%%%%
%%%%%%%%%%%%%%%%%%%%%%%%%%%%%%%%%%
\vskip 1truecm
\subsec{Coupling Gravity to Chiral Fermions}

Let's consider the theory in $D=4k+2$ dimensions. The coupling of gravitational field with chiral
fermions is given as usual by

\eqn\genact{
I_{int} = \int d^{4k+2}x \det (e) *  e^{\mu a}(x) * {1 \over 2} \overline{\psi}(x) * i
\Gamma_a
D_{\mu} 
\bigg({1 - \bar{\Gamma} \over 2} \bigg) \psi(x),}
where $D_{\mu}$ is the covariant derivative with respect to the spin connection $\omega^{ab}_{\mu}$
given by $D_{\mu} \psi(x)= \partial_{\mu} \psi(x) + {1\over 2} \omega_{\mu cd} \sigma^{cd} 
\psi(x)$, with $\sigma^{cd} = {1\over 4} [\Gamma^c,\Gamma^d]$, $\bar{\Gamma} = \Gamma_1 
\dots \Gamma_{4k+2}$ and the $\Gamma$'s are the Dirac matrices in euclidean $4k+2$
dimensions.

Our noncommutative action splits into two parts $I_{int} = I_1 + I_2$ where

\eqn\actone{
I_1= {1\over 2} \int dx \det(e) * e^{\mu a}(x) * \overline{\psi}(x) * i \Gamma_a
\buildrel{\leftrightarrow}\over {\partial}_{\mu} \bigg({1 
- \bar{\Gamma} \over 2} \bigg) \psi(x)}
and 

\eqn\acttwo{
I_2 = {1\over 4} \int dx \det(e) * e^{\mu a}(x) *  \omega_{\mu}^{ cd}(x) *i
\overline{\psi}(x)
* \Gamma_{acd} \bigg({1 - \bar{\Gamma} \over 2} \bigg) \psi(x),}
where $\Gamma_{acd}= {1\over 6}(\Gamma_a \Gamma_c \Gamma_d \pm {\rm permutations})$.

The linearization of our noncommutative action $I_{int}$ given by Eq. \genact\ leads to the
Moyal deformation of linear gravity given by the lagrangian

\eqn\lagraone{
L_1 = - {1 \over 4} i h^{\mu \nu}(x)  *\overline{\psi}(x) * \Gamma^{\mu}
\buildrel{\leftrightarrow}\over {\partial}_{\mu}
\bigg({1 - \bar{\Gamma} \over 2} \bigg) \psi(x),}
and 

\eqn\lagratwo{
L_2 = - {1\over 16} i  h_{\lambda \alpha}(x) * \partial_{\mu}  \overline{\psi}(x) * 
\Gamma^{\mu \lambda \nu} \bigg({1 - \bar{\Gamma} \over 2} \bigg) \psi(x).}

The corresponding noncommutative Feynman rules can be reading from the lagrangians \lagraone\ and
\lagratwo\ and 
they are given by

\eqn\ncfruleone{
-{i \over 4} \varepsilon^{\mu \nu} \Gamma_{\mu} \bigg({1 - \bar{\Gamma} \over 2} \bigg)
(p+p')_{\nu} 
\exp \bigg( -{i \over 2} \Theta^{\mu \nu} p_{\mu} p'_{\nu} \bigg)}
and

$$
-{i \over 16} \Gamma^{\lambda \mu \nu} \bigg({1 - \bar{\Gamma} \over 2} \bigg) 
\varepsilon^{(1)}_{\nu \alpha} \varepsilon^{(2)}_{\lambda \alpha} 
\exp \big({i \over 2}
\Theta^{\rho \sigma}
p_{\rho}p'_{\sigma} \big) 
$$

\eqn\ncfruletwo{
\times \bigg[k_{1 \mu} \exp \big({i \over 2}
\Theta^{\rho \sigma} 
k_{1 \rho}k_{2 \sigma} \big) -  k_{2 \mu} \exp \big({i \over 2}
\Theta^{\rho \sigma}
k_{1 \rho}k_{2 \sigma} \big)
\bigg],}
where $\varepsilon^{(i)}_{\mu \alpha}$ are the polarization tensors of the graviton
field.

\vskip 1truecm
%%%%%%%%%%%%%%%%%%%%%%%%%%%%%%%%%%%%%%%%%%%%%%%%%%%%%%%%%%%%%%%%
%%%%%%%%%%%%%%%%%%%%%%%%%%%%%%%%%%%%%%%%%%%%%%%%%%%%%%%%%%%%%%%%
%%%%%%%%%%%%%%%%%%%%%%%%%%%%%%%%%%%%%%%%%%%%%%%%%%%%%%%%%%%%%%%%

\newsec{Noncommutative Delbourgo-Salam Gravitational Anomaly}

Gravitational anomalies in four dimensions were studied first by Delbourgo and Salam 
\salam\ as a gravitational correction to the violation of a global symmetry responsible of
the decay: $\pi^0 \to \gamma \gamma$. This idea was further developed in Refs.
\refs{\freund,\delbourgo}. Here we shall discuss the noncommutative counterpart. Delbourgo
and Salam  \salam\ showed
that in addition to the fermion triangle diagram with three currents, the triangle  with one
current $J$ of a global symmetry and two energy-momentum tensors $T$ is also anomalous. The
corresponding contribution from the anomalous Ward identity is given by

\eqn\delsalam{ 
{1 \over 384 \pi^2} R_{\kappa \lambda \rho \sigma} R_{\mu \nu}^{\rho \sigma} 
{\varepsilon}^{\kappa \lambda \mu \nu}.
}
This is precisely proportional to the signature invariant $\sigma(X)$ (or the first Pontrjagin
class) which
together the Euler number $\chi(X)$ are the classical topological invariants of the smooth
spacetime manifold $X$. 

Now we will discuss in detail the derivation of the noncommutative counterpart of Eq. \delsalam. 
The scattering amplitude of the process in 4 dimensions is given by

$$ 
{\rm Tr} \int d^{4} p \{ \Gamma \cdot p, \Gamma_{\kappa \lambda \mu \nu} \}
{\exp \big(-{i \over 2} \Theta^{\rho \sigma} (p-k_2)_{\rho}(p+k_1)_{\sigma} \big) 
\over {[\Gamma \cdot (p+k_1)-M]}}
\varepsilon_{{\rho}_1 \sigma_1} p^{\rho_1}\Gamma^{\sigma_1} 
$$
\eqn\treeamp{
\times {\exp \big(-{i \over 2}
\Theta^{\rho \sigma} (p+k_1)_{\rho} p_{\sigma} \big) \over {(\Gamma \cdot p - M)}}
\varepsilon_{{\rho}_2 \sigma_2}p^{\rho_2}\Gamma^{\sigma_2} {\exp \big(-{i \over 2}
\Theta^{\rho \sigma}
p_{\rho}(p-k_2)_{\sigma} \big) \over {[\Gamma \cdot (p- k_2) -M]}},	
}
where we have used the Feynman rule \ncfruleone\ in each vertex of the triangle diagram 
and the
corresponding fermion propagators.

In order to evaluate this amplitude we promote the integral from 4 to $2\ell$
dimensions 

$$ 
\int d^{2\ell} p 
{(\Gamma \cdot (p+k_1) +M) \over \big[(p+k_1)^2 - M^2\big]} \cdot \exp \big(-{i \over 2}
\Theta^{\rho \sigma} (p-k_2)_{\rho}(p+k_1)_{\sigma} \big) 
\varepsilon_{{\rho}_1 \sigma_1} p^{\rho_1} \Gamma^{\sigma_1} 
{(\Gamma \cdot p +M) \over \big[p^2 - M^2\big]} 
$$
\eqn\compuno{
\times \exp \big(-{i \over 2}
\Theta^{\rho \sigma} (p+k_1)_{\rho} p_{\sigma} \big)
{(\Gamma \cdot (p-k_2) +M) \over \big[(p-k_2)^2 - M^2\big]}
\varepsilon_{{\rho}_2 \sigma_2}p^{\rho_2}\Gamma^{\sigma_2} \cdot \exp \big(-{i \over 2}
\Theta^{\rho \sigma}
p_{\rho}(p-k_2)_{\sigma} \big)	
}
and as usual we introduce the Feynman's parameters

\eqn\feynmanp{
{1 \over ABC} \equiv 2 \int_0^1 dx  \int_0^1 dy  \int_0^1 dz {\delta (1-x-y-z) \over (xA
+yB+zC)^3},
}
with $A=(p+k_1)^2 -M^2$, $B=(p-k_2)^2 -M^2$ and $C=p^2 -M^2$. After a
the redefinition of the
momenta $p \to p'= p + k_1 x - k_2 y$ we
find that $xA +yB+(1-x-y)C = p'^2 + k_3^2 xy - M^2$ and omitting the trace operation of
Dirac matrices we find that $xA +yB+(1-x-y)C = p'^2 + k_3^2 xy - M^2$ and

$$
 \int_0^1 dx dy dz \  \delta (1-x-y-z) \int {d^{2\ell}p \over (p^2 + k_3^2xy
-M^2)^3} 
$$
$$
\times {\rm Tr} \bigg\{  \{
\Gamma \cdot p, \Gamma_{\kappa \lambda
\mu \nu} \} \big[\Gamma \cdot (p+zk_1-xk_3) +M
\big]{(p+xk_2)}^{\rho_1}\Gamma^{\sigma_1}
$$
\eqn\ampli{
\times  \big[\Gamma \cdot (p+xk_2-yk_1)
+M\big]{(p-yk_1)}^{\rho_2}\Gamma^{\sigma_2} \big[\Gamma \ (p+yk_3-zk_2) +M\big]  \bigg\}
\exp \bigg(-{i
\over 2} \Theta^{\rho \sigma} k_{1 \rho}k_{2 \sigma} \bigg).
}
Here we have redefined once again $p' \to p$ and we have performed the sum of the
contributions of the phases in the
noncommutative parameter $\Theta$. Integrating out the variable $z$ and keeping only
the divergent terms we finally get

$$
2 k_2^{\rho_1} k_1^{\rho_2} \int_0^1 dx dy \ \theta (1-x-y)xy \int {d^{2\ell}p
\over (p^2 +
k_3^2xy -m^2)^3} 
$$
\eqn\seg{
\times {\rm Tr} ( \{\Gamma \cdot p, \Gamma_{\kappa \lambda \mu \nu} \} \Gamma \cdot p
\ \Gamma^{\sigma_1}\Gamma^{\sigma_2} \Gamma \cdot k_1 \Gamma \cdot k_2 ) \exp \bigg(-{i
\over 2} \Theta^{\rho \sigma} k_{1 \rho}k_{2 \sigma} \bigg),
}
where $\theta(x)$ is the Heavside function. Now we proceed to compute the trace of
products of gamma matrices using the cyclic property of the trace by using the
identity ${\rm Tr} \big( \Gamma_{\kappa \lambda \mu \nu} \Gamma^{\sigma_1}\Gamma^{\sigma_2}
\Gamma^{\alpha} \Gamma^{\beta} \big) = 2^{\ell}
\delta^{[\sigma_1}_{[\kappa}\delta^{\sigma_2}_{\lambda} \delta^{\alpha}_{\mu}
\delta^{\beta]}_{\nu]} = 2^{\ell} \varepsilon^{\sigma_1 \sigma_2 \alpha \beta}
\varepsilon_{\kappa \lambda \mu
\nu}$ we finally obtain 

$$
2 k_2^{\rho_1} k_1^{\rho_2} \int_0^1 dx dy 
\ \theta (1-x-y)xy  \int {d^{2\ell} p \over (p^2 + k_3^2xy -m^2)^3} 
2 {\bigg({\ell-2\over \ell} \bigg)}
$$
\eqn\finalone{
\times {\rm Tr} \bigg[ \{ \Gamma \cdot p, \Gamma_{\kappa \lambda \mu \nu} \}
\Gamma^{\sigma_1}\Gamma^{\sigma_2} \Gamma \cdot k_1 \Gamma \cdot k_2 + \dots \bigg] 
\exp \bigg(-{i
\over 2} \Theta^{\rho \sigma} k_{1 \rho}k_{2 \sigma} \bigg)
}
or in other form

$$
2^{\ell +1} k_2^{\rho_1} k_1^{\rho_2} \int_0^1 dx dy
\ \theta (1-x-y)xy  \int {d^{2l} p \over (p^2 + k_3^2xy -m^2)^3}
2 {\bigg({\ell-2\over \ell} \bigg)}
$$
\eqn\finaldos{
\times k_{1 \alpha} k_{2 \beta} \varepsilon^{\sigma_1 \sigma_2 \alpha \beta}
\varepsilon_{\kappa \lambda \mu \nu}  \exp \bigg(-{i
\over 2} \Theta^{\rho \sigma} k_{1 \rho}k_{2 \sigma} \bigg).
}
In the integration on the momentum $p$ we have used the following identity

\eqn\ident{
\int d^{2 \ell}p {p^2 \over (p^2 + k_3^2 xy - M^2)^3} = {i \pi^{\ell} \over
(k_3^2 xy
-M^2)^{3-\ell}} {\Gamma (2 - \ell) \over \Gamma(3)} \ell (k_3^2 xy - M^2).
}
Then finally we obtain
$$
2^{\ell+1} (\ell-2 )k_2^{\rho_1}
k_1^{\rho_2} \varepsilon^{\sigma_1 \sigma_2 \alpha \beta}
\varepsilon_{\kappa \lambda \alpha \beta} k_{1\alpha}k_{2\beta}
\exp \bigg(-{i
\over 2} \Theta^{\rho \sigma} k_{1 \rho}k_{2 \sigma} \bigg)
$$
\eqn\last{
\times {(4\pi)}^{-\ell}\Gamma(2-\ell) \int 
{(k_3^2 xy -M^2)}^{\ell-2} i xy \ \theta (1-x-y) dx
dy + \dots.
}

Performing the expansion of the gamma function $\Gamma(\varepsilon)$ for small values of
$\varepsilon$ with $\varepsilon = 2 - \ell$, taking the limit $\ell \to 2$
and evaluating the integral in $x$ and $y$ we finally get

\eqn\fourier{
-i { k_2^{\rho_1} k_1^{\rho_2}  \over 12 \pi^2 }
\varepsilon^{\sigma_1 \sigma_2 \alpha \beta}
\varepsilon_{\kappa \lambda \alpha \beta} k_{1\alpha}k_{2\beta} \exp \big(-{i
\over 2} \Theta^{\rho \sigma} k_{1 \rho}k_{2 \sigma} \big).
}

Taking into account the most general Lorentz invariant amplitude we get

\eqn\general{
-{i \over 192 \pi^2} 
\varepsilon_{\rho_1 \sigma_1}(k_1)\varepsilon_{\rho_2 \sigma_2}(k_2)
 k_{1\alpha}k_{2\beta} \varepsilon^{\sigma_1 \sigma_2 \alpha \beta}
\big(\eta^{\rho_1 \rho_2} k_1 \cdot k_2- k_1^{\rho_2}k_2^{\rho_1} \big)
\varepsilon_{\kappa \lambda \alpha \beta} \exp \bigg(-{i
\over 2} \Theta^{\rho \sigma} k_{2 \rho} k_{1 \sigma}\bigg).
}

In the coordinate space this expression can be rewritten as

\eqn\coord{
\varepsilon^{\sigma_1 \sigma_2 \alpha \beta} \bigg( \partial_{\alpha} \partial_{\gamma}
h_{\rho_1 \sigma_1} * \partial_{\beta} \partial^{\gamma}
h^{\rho_1}_{\sigma_2} -  \partial_{\alpha} \partial^{\rho_2}
h_{\rho_1 \sigma_1} * \partial_{\beta} \partial^{\rho_1}
h_{\rho_2 \sigma_2}\bigg)\varepsilon_{\kappa \lambda \alpha \beta}.
}
This equation can be rewritten as

\eqn\signature{
{1 \over 384 \pi^2} R_{\kappa \lambda \rho \sigma} * R_{\mu \nu}^{\rho \sigma} 
{\varepsilon}^{\kappa \lambda \mu \nu}.
}
This is precisely the noncommutative signature invariant $\widehat{\tau}(X)= \int R *
\widetilde{R} \ 
d^4x$, where the tilde over $R$ stands for the Hodge dual with respect the tangent space
indices. Compare this with the noncommutative signature $\widehat{\sigma}(X)$ of Ref.
\nctopo\ where the Hodge duality was associated to the tetrad indices.

\vskip 1truecm
%%%%%%%%%%%%%%%%%%%%%%%%%%%%%%%%%%%%%%%%%%%%%%%%%%%%%%%%%%%%%%%%
%%%%%%%%%%%%%%%%%%%%%%%%%%%%%%%%%%%%%%%%%%%%%%%%%%%%%%%%%%%%%%%%
%%%%%%%%%%%%%%%%%%%%%%%%%%%%%%%%%%%%%%%%%%%%%%%%%%%%%%%%%%%%%%%%

\newsec{Noncommutative Pure Gravitational Anomaly in Two Dimensions}

In the previous section we have introduced Feynman rules for noncommutative
perturbative quantum gravity relevant to compute the chiral gravitational anomalies. Before
to compute the noncommutative gravitational anomaly in $D=4k+2$ dimensions in this
section we are going into the details of the computation of the pure gravitational anomaly in
two dimensions. We will follow the notation and conventions of Ref. \alwitt.

In two dimensions the noncommutative action for a Majorana-Weyl fermion in a gravitational field is
given by $I= \int d^2 x \ det(e) * e^{\mu a}(x)* {1 \over 2} \bar{\psi}(x) * i
\Gamma_a \partial_{\mu}\psi(x).$ At the linearized level, the lagrangian is given by

\eqn\one{
L_{int} = -{1 \over 4}  h^{\mu \nu}(x) *  i \bar{\psi}(x) * \Gamma_{\mu} \partial_{\nu}
\psi(x).
}
The corresponding energy-momentum tensor is given by

\eqn\two{
T_{\mu \nu}(x) = {1 \over 2} i \bar{\psi}(x) * \Gamma_{\mu} \partial_{\nu}\psi(x).
}

In the light-cone coordinates $x^{\pm} ={1 \over \sqrt{2}}(x^0 \pm x^1)$, Dirac matrices
are decomposed into  $\Gamma^{\pm} ={1 \over \sqrt{2}} (\Gamma^0 + \Gamma^1)$, with
$(\Gamma^{\pm})^2 =0$ and $\Gamma^+ \Gamma^- + \Gamma^- \Gamma^+ =2$. In these
coordinates
the energy-momentum tensor is given by

\eqn\three{
T_{++}(x) = {1 \over 2} i \bar{\psi}(x) * \Gamma_{+} \partial_{+} \psi(x),
}
while the interaction action \one\ of the gravitational field with fermions in
the light-cone coordinates reduces to

\eqn\four{
L_{int} = -  {1 \over 4} i h_{--}(x) * \bar{\psi}(x) * \Gamma_{+}
\partial_{+} \psi(x),
}
then only the component $h_{--}(x)$ of the graviton is coupled to chiral matter described by the
the component $T_{++}(x)$ of the energy-momentum tensor. The effective action to second order in
the metric perturbation $h$ is encoded in the
two-point correlation function

\eqn\five{
U(p)= \int d^2 x \exp \big( i p \cdot x \big) \langle \Omega | T(T_{++}(x)* T_{++}(0)|
\Omega \rangle,
}
where 
$$
\langle \Omega | T(T_{++}(x)*T_{++}(0)| 
\Omega \rangle = - {1 \over 4} \int \prod_{i=1}^2 {d^2q_i \over (2 \pi)^2}  \prod_{j=1}^2
{d^2q'_i \over (2 \pi)^2} \langle \widetilde{\bar{\psi}}(q_1) \gamma_+ \partial_+
{\psi}(q_1') \cdot \widetilde{\bar{\psi}}(q_2) \gamma_+ \partial_+ 
{\psi}(q_2') \rangle 
$$
\eqn\six{
\times \exp \big( i(q_1 - q_1')x \big) \exp \big( i(q_2 - q_2')x \big)
\exp \bigg( i{\Theta_{\rho \sigma} \over 2} \sum_{j=1}^3 q_j^{\rho} q'^{\sigma}_j \bigg).
}

The naive Ward identity is given by $p_- U(p) =0$. This should imply $U(p) =0$ for all $p_-$,
thus it should be an anomaly. Thus we can compute $U(p)$ by evaluating the corresponding one-loop
diagram with two external gravitons, this yields

$$
U(p) = - {1 \over 4} \int {d k_+ dk_- \over (2 \pi)^2} (2k + p)_+^2 {k_+ \exp \big(-{i\over
2} \Theta^{\rho \sigma} k_{\rho} p_{\sigma} \big)\over k_+ k_- + i
\varepsilon}
$$
$$
\times {(k+p)_+ \exp \big(-{i\over 2}
\Theta^{\rho \sigma} p'_{\rho} k_{\sigma} \big)\over (k+p)_+(k+p)_- + i \varepsilon} 
\delta (p+p') \cdot \exp \big(i(p+p')x \big)
$$
\eqn\seven{
=- {1 \over 4} \int {d k_+ dk_- \over (2 \pi)^2} (2k + p)_+^2 {1 \over k_- + i
\varepsilon /k_+} {\exp \big(-{i\over 2}
\Theta^{\rho \sigma} p'_{\rho} p_{\sigma} \big) \over (k+p)_- + i \varepsilon / (k+p)_+}
\cdot \delta (p+p') \cdot \exp \big(i(p+p')x \big),
}
where we have used the Feynman rule \ncfruleone\ to compute $U(p)$.

In light-cone coordinates the Moyal product is given by $\exp \big(-{i\over 2}
\Theta^{\rho \sigma} p'_{\rho} p_{\sigma} \big)$ $= \exp \big( -{1 \over 2}
\Theta^{+-}(p'_+
p_- -  p'_- p_+) \big)$. Thus by analytic methods the computation of the integrals gives

$$
U(p) = {i\over 8 \pi} \int_{-p_+}^0 dk_+ {(2k +p)_+^2 \over p_-} \exp \big(-{i\over 2}
\Theta^{\rho \sigma} p'_{\rho} p_{\sigma} \big) \delta(p+p')
$$
\eqn\eight{ 
= {i\over 24 \pi}{p_+^3 \over p_-} \exp \bigg(-{i\over 2}
\Theta^{\rho \sigma} p'_{\rho} p_{\sigma} \bigg) \exp \big(i(p+p')x \big) \delta(p+p').
}
Thus the anomalus gravitational Ward identity is given by

\eqn\award{
p_- U(p) = {i\over 24 \pi}{p_+^3} \exp \bigg(-{i\over 2} \Theta^{\rho \sigma} p'_{\rho}
p_{\sigma}
\bigg) \exp \big(i(p+p')x \big) \delta(p+p').
}

The computation of the two-graviton diagram coupled with chiral fermions in the
noncommutative theory is given by the effective action

$$
L^{eff}_+(h_{\mu \nu}) = - {1 \over 192 \pi} \int d^2 p d^2 p'{p_+^3 \over p_-} h_{--}(p)
$$
\eqn\effact{
\times \exp \bigg(-{i \over 2}
\Theta^{\rho \sigma} p'_{\rho} p_{\sigma} \bigg)
h_{--}(p') \exp \big(i(p+p')x \big) \delta(p+p').
}
Similarly to the usual commutative case there is no way to add generic counterterms
$\Delta L^{eff}_+$ such that $ L^{eff}_+ + \Delta L^{eff}_+$be invariant under
general coordinate transformations.

Thus, let us consider a Dirac fermion in $1+1$ dimensions, then we have the corresponding
action $L^{eff}_D$ is the
superposition of $L^{eff}_+$ and its corresponding parity conjugate  $L^{eff}_-$ resulting 

$$
L^{eff}_D(h_{\mu \nu}) = - {1 \over 192 \pi} \int d^2 p d^2 p'\bigg[ {p_+^3 \over p_-}
h_{--}(p)
\exp \bigg(-{i \over 2}\Theta^{\rho \sigma} p'_{\rho} p_{\sigma} \bigg) h_{--}(p')
$$
\eqn\nine{
+  {p_-^3 \over p_+} h_{++}(p) \exp \bigg(-{i\over 2}
\Theta^{\rho \sigma} p'_{\rho} p_{\sigma} \bigg) h_{++}(p') \bigg] \exp \big(i(p+p')x \big)
\delta(p+p').
}
This action is not invariant under infinitesimal general coordinate
transformations $\delta x^{\mu} =
\varepsilon^{\mu}$, $h_{\mu \nu}$ transforms as $ \delta h_{\mu \nu}(x) = - \partial_{\mu}
\varepsilon_{\nu}(x) - \partial_{\nu} \varepsilon_{\mu}(x)$ or in the momentum space

\eqn\trasfdiff{
\delta h_{++}(p) = -2 i p_+ \varepsilon_+, \ \ \ \ \  \delta h_{+-}(p) = -i p_- \varepsilon_+
- i p_+ \varepsilon_-, \ \ \  \delta h_{--}(p) = -2 i p_- \varepsilon_-.
}
However in this case there exist a counterterm $\Delta L^{eff}_D$ which can be added to
$L^{eff}_D$ such that it become invariant under general coordinate transformations

$$
\Delta L^{eff}_D = - {1 \over 192 \pi} \int d^2 p d^2 p' \bigg[ {p_+^3 \over p_-}
h_{--}(p)
\exp \bigg(-{i\over 2}
\Theta^{\rho \sigma} p'_{\rho} p_{\sigma} \bigg)h_{--}(p')
$$

$$
+  {p_-^3 \over p_+} h_{++}(p)  \exp \bigg(-{i\over 2}
\Theta^{\rho \sigma} p'_{\rho} p_{\sigma} \bigg) h_{++}(p')
+ 2 p_+p_- h_{++}(p) \exp \bigg(-{i\over 2}
\Theta^{\rho \sigma} p'_{\rho} p_{\sigma} \bigg) h_{--}(p')  
$$
$$
- 4 p_+^2 h_{--}(p)
\exp \bigg(-{i\over 2}
\Theta^{\rho \sigma} p'_{\rho} p_{\sigma} \bigg) h_{+-}(p')
- 4 p_-^2
h_{++}(p) \exp \bigg(-{i\over 2}
\Theta^{\rho \sigma} p'_{\rho} p_{\sigma} \bigg)h_{+-}(p')  
$$
\eqn\lasteff{
+ 4 p_+p_- h_{+-}(p) \exp \bigg(-{i\over 2}
\Theta^{\rho \sigma} p'_{\rho} p_{\sigma} \bigg) h_{+-}(p') \bigg] \delta(p+p').
}

It is easy to see that this action can be rewritten in a compact form as following

\eqn\ayya{
\Delta L^{eff}_D  = - {1 \over 192 \pi} \int d^2 p d^2 p'{ R(p) \exp \big(-{i\over 2}
\Theta^{\rho \sigma} p'_{\rho} p_{\sigma} \big) R(p') \over p_+ p_-} \delta(p+p'),
}
which after integration in the variable $p'$ gives the usual correction to the commutative
counterpart of \lasteff. 
\eqn\countert{
\Delta L^{eff}_D= - {1 \over 192 \pi}
\int d^2 p { R(p)  R(-p) \over p_+ p_-}.
}
where $R(p)$ is the linearized term of the noncommutative curvature scalar
which is given by $R(p) = p_+^2 h_{--} + p_-^2 h_{++} - 2 p_+ p_- h_{+-}$.

There a quantum correction to $T_{+-}(p)=0$ which holds classically due to the
introduction of $h_{+-}$ in the counterterm lagrangian
$\Delta L^{eff}_D$ and we have an expectation value of $T_{+-}$ different to zero which
gives rise to the
gravitational anomaly 

\eqn\anomaly{ 
\langle 2 T_{+-}(p) \rangle = -2 {\delta \Delta L^{eff}_D \over \delta h_{+-}(-p)} = -{1 \over 24
\pi} R(p).
}
By momentum conservation we have in the above analysis that $p'=-p$ through the
$\delta(p+p')$ and the phase 
factor $\exp \big(-i
\Theta^{\rho \sigma} p'_{\rho} p_{\sigma} \big)$ is equal to one and therefore there
is no a modification to the gravitational anomaly in two dimensions in a noncommutative 
space.

%%%%%%%%%%%%%%%%%%%%%%%%%%%%%%%%%%%%%%%%%%%%%%%%%%%%%%%%%%%%%%%%
%%%%%%%%%%%%%%%%%%%%%%%%%%%%%%%%%%%%%%%%%%%%%%%%%%%%%%%%%%%%%%%%
\newsec{Noncommutative Gravitational Anomalies in $D=4k+2$ dimensions}

%%%%%%%%%%%%%%%%%%%
%%%%%%%%%%%%%%%%%%%
\subsec{Preliminaries}

In this subsection we compute the one-loop diagram of ${D\over 2}+1 =2k+2$ external
gravitons of
momentum
$p_{\mu}^{(i)}$ and polarizations $\varepsilon_{\mu \nu}^{(i)}$ with $i= 1, \dots ,2k+2$.
In
this
computation we will follow Ref. \alwitt\ by using Adler's prescription of an equivalent
diagram with $2k+1$ external gravitons with only one insertion of an axial factor ${1\over
2} (1- \bar{\Gamma})$ and $2k+1$ non-anomalous vertices. In what follows we assume that
the polarization tensor
$\varepsilon_{\mu \nu}$ given by $ \varepsilon_{\mu \nu} = i (p_{\mu}
\varepsilon_{\nu} +
p_{\nu} \varepsilon_{\mu})$ (where $\varepsilon_{\mu}$ is the parameter involved in the
coordinate transformation $x^{\mu} \to x^{\mu} + \varepsilon^{\mu}$)
will be factorized as:
$\varepsilon^{(i)}_{\mu \nu} = \varepsilon^{(i)}_{\mu} \cdot
\varepsilon^{(i)}_{\nu}$.

Then the total one-loop amplitude is proportional to:

$$
{\cal A} \propto {\rm Tr} \bigg[ \bar{\Gamma} \exp \bigg(-{i
\over 2} \Theta^{\rho \sigma} k_{\rho}(k - p^{(1)} - \dots -
p^{(2k+1)})_{\sigma} \bigg) (\not \! \! k +M) 
$$
$$
\times \not \! \! \varepsilon^{(1)} 
\exp \bigg(-{i \over 2} \Theta^{\rho \sigma} (k - p^{(1)})_{\rho}k_{\sigma} \bigg)(\not \!
\! k 
- \not 
\! \! p^{(1)} +M) \not \! \! \varepsilon^{(2)} \exp \bigg(-{i \over 2} \Theta^{\rho
\sigma} (k
- p^{(1)} - p^{(2)})_{\rho}(k - p^{(1)})_{\sigma} \bigg)
$$  
$$
\times(\not \! \! k - \not
\! \! p^{(1)} - \not \! \! p^{(2)}+M) \not \! \! \varepsilon^{(3)} \exp \bigg(-{i \over 2}
\Theta^{\rho \sigma} (k - p^{(1)} - p^{(2)} -
p^{(3)})_{\rho}(k - p^{(1)} - p^{(2)})_{\sigma} \bigg)  
$$
$$
\dots   \times \not \! \! 
\varepsilon^{(2k+1)} 
\exp \bigg(-{i
\over 2} \Theta^{\rho \sigma} (k - p^{(1)} - \dots - p^{(2k+1)})_{\rho}(k - p^{(1)} - \dots -
p^{(2k)})_{\sigma} \bigg)
$$
\eqn\genampl{
\times (\not \! \! k - \not
\! \! p^{(1)} - \dots - \not \! \! p^{(2k+1)}+M) \bigg],
}
where we have used the Feynman rule \ncfruleone\ in each non-anomalous vertex. In the
amplitude we have omitted a $(2k+1)$ factors of the form ${1 \over  p^2 - M^2}$ in each
non-anomalous
vertex.

Now, in order eliminate Dirac matrices we require that ${\rm Tr} \big( \bar{\Gamma}
\Gamma_{\mu_1} \Gamma_{\mu_2} \dots \Gamma_{\mu_{4k+2}} \big) = -2^{2k+1}
\varepsilon_{\mu_1 \mu_2 \dots \mu_{4k+2}}$. Thus we can factorize the dependence on
the noncommutativity parameter $\Theta$ 

\eqn\fact{
{\cal A} \propto  2^{2k+1} M R\big(\varepsilon^{(i)}, p^{(j)}\big),
}
where $R(\varepsilon^{(i)} p^{(j)})$ is a kinematical factor which depends only on the 
external momenta and polarization vectors 

\eqn\erre{
R(\varepsilon^{(i)}, p^{(j)}) = - \varepsilon_{\mu_1 \mu_2 \dots \mu_{4k+2}}
p^{(1)}_{\mu_1} 
\varepsilon^{(1)}_{\mu_2}  p^{(2)}_{\mu_3} \varepsilon^{(2)}_{\mu_4} \dots 
p^{(2k+1)}_{\mu_{4k+1}} \varepsilon^{(2k+1)}_{\mu_{4k+2}},
}
leaving $R(\varepsilon^{(i)}, p^{(j)})$ independent on $\Theta$.

Using the Feynman rule \ncfruleone\ in each one of the $2k+1$ vertices we have for the
$i$-th vertex there is the insertion of a factor: $-{1\over 4} i \varepsilon^{(i)}_{\mu}
(p+p')^{\mu}{1 \over p^2 - M^2} \exp \big(-{i \over 2} \Theta^{\rho \sigma} p_{\rho}
p'_{\sigma}\big)$, where $p$ is the 
incoming momentum and $p'$ is the outgoing momentum. The whole contribution is encoded in
the 
amplitude ${\cal Z}( \varepsilon^{(i)}, p^{(j)}, \Theta)$. The total amplitude is then
given by

\eqn\total{
I_{1\over 2} = 2^{2k+1} M^2 R(\varepsilon^{(i)}, p^{(j)}) \cdot {\cal Z}(
\varepsilon^{(i)},p^{(j)},\Theta),
}
where ${\cal Z}$ can be reinterpreted as the amplitude for a charged scalar field of mass
$M$ and charge
${1\over 4}$ in a loop coupled to $(2k+2)$ photons of momenta $p^{(j)}$ and polarization
tensors
$\varepsilon^{(i)}$ in a noncommutative spacetime.

Then all the information of the noncommutative parameter is in the amplitude ${\cal Z}$ and
we need to find a way of computing the amplitude 

\eqn\totalone{ 
{\cal Z}( \varepsilon^{(i)}, p^{(j)}, \Theta) = \int d^{4k +2}k { \prod_{j=1}^{2k+2}
 \exp \bigg(-{i \over 2} \Theta^{\rho \sigma} \sum_{j} l^{(j)}_{\rho}
l^{(j+1)}_{\sigma} \bigg) \varepsilon \cdot (l_j + l_{j+1}) \over \prod_{j=1}^{2k+2}(l_j^2
- M^2)}. } 
As in the commutative case, this problem
can be carried over to the residual problem of the computation of this amplitude
for a one-loop diagram with $2k+2$ external
photons interacting noncommutatively with a massive complex scalar field of charge ${1\over
4}$ with
propagators
$i/(p^2 - M^2)$ under the condition that in the $i$-th vertex we have a factor $-{1\over
4}i
\varepsilon^{(i)}_{\mu}(p+p')^{\mu}  \exp \big(-{i \over 2} \Theta^{\rho \sigma} p_{\rho}
p'_{\sigma}\big)$, where $p$ and $p'$ \alwitt. This problem was discussed by Schwinger 
\schw\ for the commutative case and used by \alwitt\ to compute ${\cal Z}$. In this paper
we follow the same strategy for the noncommutative case. In the next subsection, we give
the details of the explicit computation of this noncommutative residual interaction.
Basically we will have a noncommutative one of this interaction in each non-anomalous
vertex and we find an exact solution for it and then apply it to
compute ${\cal Z}$.

%%%%%%%%%%%%%%%%%%%%%%
%%%%%%%%%%%%%%%%%%%%%
\subsec{Explicit Computation of the Noncommutative Residual Interaction}

We start from a theory for a complex scalar field of mass $M$ coupled to an abelian
gauge field in a noncommutative space. Due to the noncommutative bosonization, this system
will be equivalent to a Schwinger model. The Schwinger model has been discussed in the
noncommutative context in \refs{\madoregrosse,\grosse,\bala,\ydri}, however in the present
paper we follow a different procedure. Consider the following action

\eqn\alfa{
L= \int d^{2p}x \big( D^{\mu} \bar{\phi} * D_{\mu} \phi +  M^2 \bar{\phi} * \phi \big),}
with $D_\mu\phi = \partial_\mu \phi - ie A_\mu * \phi$ and $D_\mu\bar{\phi} =
\partial_\mu \bar{\phi} +ie \bar{\phi} * A_\mu.$ If we use the definition for star
product $(f * g)(x)= f e^{\overleftarrow{\partial}_\alpha \Theta^{\alpha \beta}
\overrightarrow{\partial}_\beta} g(x),$ where $\overleftarrow{\partial}_\alpha \Theta^{\alpha
\beta} \overrightarrow{\partial}_\beta \equiv {i \over 2} \Theta^{\alpha \beta}
\overleftarrow{\partial}_\alpha  \overrightarrow{\partial}_\beta$.
Some results found by Schwinger \schw\ were used in Ref. \alwitt\ as an tool to
compute the gravitational anomaly in $4k+2$  dimensions for gravitons coupled to
spin-${1\over 2}$ fields.

The first term of the RHS of \alfa\ noncommutative action is given by

\eqn\gama{
\int d^{2p}x D^\mu \bar{\phi} * D_\mu \phi =\int \bigg( \partial^\mu \bar{\phi} + ie
\bar{\phi}e^{\overleftarrow{\partial}_\alpha \Theta^{\alpha \beta}
\overrightarrow{\partial}_\beta} A^\mu \bigg)  \bigg( \partial_\mu \phi - ie A_\mu
e^{\overleftarrow{\partial}_\alpha \Theta^{\alpha \beta} \overrightarrow{\partial}_\beta} \phi
\bigg),}
where we have used the cyclicity property of the trace $\int dx f(x) * g (x) = \int dx f(x) 
g (x)$ for any $f$ and $g$. Expanding RHS of this expression and integrating by parts, we can
factorize it as:

\eqn\deltaa{
\int \bar{\phi} \bigg\{-\bigg[ \partial^\mu-ie\bigg( e^{\overleftarrow{\partial}_\alpha
\Theta^{\alpha \beta} \overrightarrow{\partial}_\beta}A^\mu\bigg)\bigg] \bigg[\partial_\mu
-ie \bigg(A_\mu e^{\overleftarrow{\partial}_\alpha \Theta^{\alpha \beta}
\overrightarrow{\partial}_\beta}\bigg)\bigg]\bigg\}\phi.}

Then the starting action \alfa\ results:

\eqn\seta{
L=\int \bar{\phi} \bigg\{-\bigg[ \partial^\mu-ie\bigg( e^{\overleftarrow{\partial}_\alpha
\Theta^{\alpha \beta} \overrightarrow{\partial}_\beta}A^\mu\bigg)\bigg] \bigg[\partial_\mu
-ie \bigg(A_\mu e^{\overleftarrow{\partial}_\alpha \Theta^{\alpha \beta}
\overrightarrow{\partial}_\beta}\bigg)\bigg] + M^2 \bigg\}\phi.}

Let us define the euclidean partition function for a complex massive scalar field
propagating
in a constant electromagnetic field as:

\eqn\pf{
Z= \int [{\cal D}\phi]  [{\cal D}\bar{\phi}] \exp\big(-L\big),}
then the effective action $\Gamma$ is related to the partition function in the form $Z=
e^{-\Gamma}$

\eqn\teta{
\Gamma= {\rm Tr} \ln \bigg\{-\bigg[ \partial^\mu-ie\bigg( e^{\overleftarrow{\partial}_\alpha
\Theta^{\alpha \beta} \overrightarrow{\partial}_\beta}A^\mu\bigg)\bigg] \bigg[\partial_\mu
-ie \bigg(A_\mu e^{\overleftarrow{\partial}_\alpha \Theta^{\alpha \beta}
\overrightarrow{\partial}_\beta}\bigg)\bigg] + M^2 \bigg\}.}

The Schwinger representation for the logaritmic function can be expressed as follows:

\eqn\kapa{
\Gamma = - {\rm Tr} \int_0^\infty {ds\over s}\bigg\{ e^{-s\bigg[-\bigg( \partial^{\mu}-ie\big(
e^{\overleftarrow{\partial}_\alpha \Theta^{\alpha \beta}
\overrightarrow{\partial}_\beta}A^\mu\big)\bigg) \bigg(\partial_\mu -ie \big(A_\mu
e^{\overleftarrow{\partial}_\alpha \Theta^{\alpha \beta}
\overrightarrow{\partial}_\beta}\big)\bigg) + M^2 \bigg]}- e^{-s} \bigg\}.}

We can write the strength field $F_{\mu \nu}$ of the constant electromagmnetic field as
usual in the Schwinger representation, as a diagonal block matrix and
work only with a generic block of two components.  Also we will take the following gauge:

\eqn\lamdaaa{
A_1=0, \ \ \ \ \ \ \  A_2=Fx^1.}
Notice that in this gauge, higher order terms than the first order vanish in the expansion of
the Moyal product. With this in mind we can calculate the effective action $\Gamma$ as:

$$
\Gamma= -{\rm Tr}\int_0^\infty {ds\over s} e^{-sM^2} \exp \bigg\{s\bigg[ (\partial^\mu
-ieA^\mu)(\partial_\mu-ieA_\mu)  - ie {i\over 2} \Theta^{\alpha\beta}
(\partial^\mu\partial_\alpha A_\mu \partial_\beta + \partial_\alpha
A_\mu\partial^\mu\partial_\beta)
$$
$$
- ie {i\over 2}
\Theta^{\alpha\beta}(\partial_\alpha\partial_\beta A^\mu\partial_\mu + \partial_\beta A^\mu
\partial_\alpha \partial_\mu)-e^2 \bigg({i\over 2} \Theta^{\alpha\beta}( A^\mu \partial_\alpha
A_\mu \partial_\beta - \partial_\alpha\partial_\beta A^\mu A_\mu - \partial_\beta A^\mu
\partial_\alpha A_\mu)
$$

\eqn\jupsilon{
- \bigg( {i\over 2} \bigg)^2 \Theta^{\alpha\beta}
\Theta^{\lambda\delta} (\partial_\alpha\partial_\beta A^\mu\partial_\lambda
A_\mu\partial_\delta + \partial_\beta A^\mu \partial_\alpha \partial_\lambda A_\mu
\partial_\delta + \partial_\beta A^\mu\partial_\lambda A_\mu\partial_\alpha\partial_\delta )
\bigg) \bigg]\bigg\} -e^{-s}.}

Now we focus in the operator in the exponential given by 

$$\bigg[ \partial^\mu-ie\big(
e^{\overleftarrow{\partial}_\alpha \Theta^{\alpha \beta}
\overrightarrow{\partial}_\beta}A^\mu\big)\bigg] \bigg[\partial_\mu -ie \bigg(A_\mu
e^{\overleftarrow{\partial}_\alpha \Theta^{\alpha \beta}
\overrightarrow{\partial}_\beta}\bigg)\bigg],
$$
it is necessary to consider only one
two-dimensional subspace expanded by the corresponding block. After some simplifications we
obtain that this operator gives $\partial_1^2 + \partial_2^2 -ieFx^1\partial_2 + i^2e^2F^2
(x^1)^2 + e\Theta
F\partial_2^2-{i\over 2}e^2\Theta F^2x^1\partial_2 + {1\over 4}e^2\Theta^2F^2\partial_2^2$.
Thus finally it factorizes as

\eqn\fact{
\partial_1^2 + \bigg[ \big(1+{e\Theta F \over 2}\big) -ieFx^1\bigg]^2. }

Now, after substitute $\hat{p}_j=-i\partial_j$ in this last expression we get

\eqn\oper{
-\bigg\{ \hat{p}_1^2 +\bigg[\big(1+{e\Theta F\over 2}\big)\hat{p}_2
-eF\hat{x}^1\bigg]^2\bigg\}.}

Then, in order to get the efective action \kapa\ we need to compute alternatively 

\eqn\inte{
I= {\rm Tr} \exp\bigg(-s\bigg\{ \hat{p}_1^2 +\bigg[\big(1+{e\Theta F\over 2}\big)\hat{p}_2
-eF\hat{x}^1\bigg]^2\bigg\}\bigg)}
Considering the problem in a box of volume $L\times L$ and using the definition of the trace we
finally get

\eqn\boxxer{
I= \bigg(\int dx_2\int {dp_2 \over 2\pi} \bigg) {\rm Tr_1} \exp\bigg(-s\bigg\{ \hat{p}_1^2
+e^2F^2\bigg[\hat{x}^1 -\bigg({1\over eF}+ {\Theta\over 2}\bigg)p_2\bigg]^2\bigg\}\bigg).}

Thus, we obtain the effective action of a one-dimensional harmonic oscillator 
Then $\big({1\over eF}+ {\Theta  \over 2}\big)p_2$ with effective center at $(x_1)_0=
\big({1\over eF}+ {\Theta\over 2}\big)p_2$. Boundary condition: $0\leq p_2 \leq \big({1
\over
eF}+ {\Theta\over 2}\big)^{-1}L$ implies that

\eqn\casi{
I=(VolR^2){1\over 2\pi}\bigg({eF\over 1+{\Theta e F\over 2}}\bigg) {\rm tr}_y\exp \{-s
(\hat{p}_y^2+e^2F^2\hat{y}^2)\}}
where $L=VolR$. This trace yields precisely the partition function of an ordinary harmonic
oscillator given by

\eqn\osilador{
{\rm Tr}_y e^{-s (\hat{p}_y^2+e^2F^2\hat{y}^2)} = {1\over 2}{1\over \sinh(seF)}.}
We finally obtain 

\eqn\final{
I= (VolR^2){1\over 4\pi}\bigg({eF\over 1+{\Theta e F\over 2}}\bigg) {1\over \sinh(seF)}.}
Thus the effective action \kapa\ is given by

\eqn\last{
\Gamma \propto - \int_0^{\infty} {ds \over s} \prod_{j=1}^p {1 \over 4\pi} \bigg({x_j/2\over
1+{\Theta x_j\over 4}}\bigg)
 {1\over \sinh({sx_j\over 2})} e^{-sM^2} + \ {\rm constant},}
where $x_j = 2eF$.

%%%%%%%%%%%%%%%%%%%%%%%%%%%%%%%%
%%%%%%%%%%%%%%%%%%%%%%%%%%%%%%%%

\subsec{Gravitational Anomaly for Spin-${1\over 2}$ Fields}

Using Eq. \last\ concerning the total amplitude of the residual interaction we get

\eqn\seta{
{\cal Z} = - \int_0^{\infty} {ds \over s} \prod_{j=1}^{2k+1} {1 \over 4 \pi}\bigg( {{1\over
2} x_j \over  \sinh ({s x_j \over 2})} \bigg)\bigg({1\over 1 + \Theta{x_j \over 4}}\bigg)
\exp(-sM^2).
}
After $s$-integration we finally get

\eqn\setauno{
{\cal Z}= - {1 \over (4 \pi)^{2k+1}} {1 \over M^2} \prod_{j=1}^{2k+1} {{1\over 2} x_j \over
4 \pi \sinh ({1 \over 2}x_j )} \bigg({1\over 1 + \Theta{x_j \over 4}}\bigg).
}
This equation also can be written as

\eqn\setados{
I_{1\over 2} = -i {1 \over (2 \pi)^{2k+1}}  R(\varepsilon^{(i)}, p^{(j)})
\widehat{A}_{\Theta}(X),
}
where

\eqn\roof{
\widehat{A}_{\Theta}(X)= \prod_{j=1}^{2k+1} \bigg({{1\over 2} x_j \over
\sinh ({1 \over 2}x_j )}\bigg) \bigg({1\over 1 + \Theta{x_j \over 4}}\bigg),
}
is the noncommutative roof-genus. Roof-genus enters in the Atiyah-Singer theorem, thus
our computation leads evidently to a noncommutative continuous deformation of
the Atiyah-Singer theorem. 

%%%%%%%%%%%%%%%%%%%
%%%%%%%%%%%%%%%%%%%
\subsec{Gravitational Anomaly for Spin-${3\over 2}$ Fields}

Now, we would like to compute the one-loop diagram of $2k+2$ external gravitons of
momentum $p_{\mu}^{(i)}$ and polarizations $\varepsilon_{\mu \nu}^{(i)}$ with $i= 1, \dots
,2k+2$ coupled to Rarita-Schwinger fields of spin  ${3 \over 2}$. In order to make
this 
computation we will use also Adler's prescription to find 
an equivalent diagram with $2k+1$ external gravitons with only one insertion of an axial
factor ${1\over 2} (1- \bar{\Gamma})$ and $2k+1$ non-anomalous vertices. 

We start from a gauge fixed linearized noncommutative contributions 

\eqn\raritauno{
L^{RS}_1 = {1 \over 4} i h^{\alpha \beta} * \psi_{\mu} * \Gamma_{\alpha}
\buildrel{\leftrightarrow}\over {\partial} \big({1 -
\bar{\Gamma}\over 2}\big) \psi^{\mu}
}
and 
\eqn\raritados{
L^{RS}_2 = {1 \over 2}i G_{\sigma \alpha \nu} * \bar{\psi}^{\sigma} * \Gamma^{\nu}
\psi^{\alpha},
}
where $G_{\sigma \alpha \nu} = \big(\partial_{\sigma} h_{\alpha \nu} - \partial_{\alpha}
h_{\sigma \nu} \big)$. The analisis of gauge fixing involves the existence of ghost fields
that modify the total amplitude and it is modified by $I_{3\over 2}(total) = I_{3\over
2}(gravitino) - I_{1\over 2}$.

Using the Feynman rules associated to $L^{RS}_1$ and $L^{RS}_2$ in each one of the
$2k+1$ vertices we have for the
$i$-th vertex there is the insertion of a factor: $-{1\over 4} i \varepsilon^{(i)}_{\mu}
(p+p')^{\mu}{1 \over p^2 - M^2} \exp \big(-{i \over 2} \Theta^{\rho \sigma} p_{\rho}
p'_{\sigma}\big)$, where $p$ is the 
incoming momentum and $p'$ is the outgoing momentum. The whole contribution is encoded in
the 
amplitude $\widehat{\cal Z}( \varepsilon^{(i)}, p^{(j)}, \Theta)$. The total amplitude is
then
given by

\eqn\totala{
I_{3\over 2} = 2^{2k+1} i M^2 R(\varepsilon^{(i)}, p^{(j)}) \cdot \widetilde{\cal Z}(
\varepsilon^{(i)},p^{(j)},\Theta),
}
where $R(\varepsilon^{(i)} p^{(j)})$ is the same kinematical factor \erre,which depends
only on the
external momenta and polarization vectors
and $\widetilde{\cal Z}$ can be regarded  as the
amplitude for the coupling of a charged (complex) noncommutative abelian vector field 
in a loop coupled to $2k+2$ photons of momenta $p^{(j)}$ and polarization
tensors $\varepsilon^{(i)}$ in a noncommutative spacetime. This noncommutative residual
interaction is described by the corresponding interaction lagrangians

\eqn\resone{
L^{(res)}_1 = {1 \over 4} A^{\mu} * \bar{\phi}_{\sigma} * \partial_{\mu}
\phi^{\sigma}
}
and 
\eqn\resdos{
L^{(res)}_2 = {1 \over 2} G_{\mu \nu} * \bar{\phi}_{\sigma}*\phi^{\sigma},
}
where $G_{\mu \nu} = \partial_{\mu}A_{\nu} - \partial_{\nu}A_{\mu} -{i\over
4}[A^{\mu},A^{\nu}]_*$. The first action \resone\ gives precisely the interaction we
discussed
in the previous subsection and which consist of $D$ complex scalars with charge ${1\over
4}$ coupled to $2k+2$ noncommutative `photons'. The second lagrangian \resdos\ corresponds
to a noncommutative magnetic moment which has the usual term $\int d^D x \
\bar{\phi}^{\mu} *
(\partial_{\mu}A_{\nu} - \partial_{\nu}A_{\mu}) * {\phi}^{\nu}$ plus an additional term
of the form

\eqn\intterm{
-{i \over 4} \int d^{4k+2} x \ \bar{\phi}^{\mu} * [A_{\mu},A_{\nu}]_* * \phi^{\nu},
}
which comes from the quadratic term of the definition of $G_{\mu \nu}$. The linear term
is
exactly the same as the spin-${1\over 2}$ case thus both terms can be gatthered and
corresponds with the computation of ${\cal Z}$ of the spin-${1\over 2}$ case in the
previous subsection. Thus in this case the only difference lies in the interaction term 
\intterm. We now proceed to compute this term. We use the cyclicity property of the
trace in order to remove the $*$-product arising in the noncommutative commutator
$[A_{\mu},A_{\nu}]_*$. With this in mind we get

$$
-{i \over 4} \int d^{4k+2} x \ \bigg\{ \bigg(\bar{\phi}_{\mu}
e^{\overleftarrow{\partial}_\alpha
\Theta^{\alpha \beta}
\overrightarrow{\partial}_\beta} A^{\mu}\bigg) \cdot \bigg(A^{\nu}
e^{\overleftarrow{\partial}_\alpha \Theta^{\alpha \beta}
\overrightarrow{\partial}_\beta} \phi_{\nu} \bigg)
$$
\eqn\modifintterm{
 -  \bigg(\bar{\phi}_{\mu}
e^{\overleftarrow{\partial}_\alpha
\Theta^{\alpha \beta}
\overrightarrow{\partial}_\beta} A^{\nu}\bigg) \cdot \bigg(A^{\mu}          
e^{\overleftarrow{\partial}_\alpha \Theta^{\alpha \beta}
\overrightarrow{\partial}_\beta} \phi_{\nu} \bigg) \bigg\}.
}
Now, using the gauge \lamdaaa\ the only term that contributes is that of second order in
$\Theta$ in equation \modifintterm. Reordering all terms and integrating by parts all
terms cancel identically, which means that the quadratic term does not contributes to the
amplitude. Then the remaining lagrangian is given by

$$
L=\int d^{4k+2}x \ \bar{\phi}^{\sigma} * \bigg\{-\bigg[ \partial^\mu-ie\bigg(
e^{\overleftarrow{\partial}_\alpha  
\Theta^{\alpha \beta} \overrightarrow{\partial}_\beta}A^\mu\bigg)\bigg] 
$$
\eqn\setaaa{
\times \bigg[\partial_\mu
-ie \bigg(A_\mu e^{\overleftarrow{\partial}_\alpha \Theta^{\alpha \beta}
\overrightarrow{\partial}_\beta}\bigg)\bigg] + M^2 - {i \over 2} F_{\mu \nu}
\bigg\}* \phi_{\sigma}.
}
Then, the effective action reads

$$
\widetilde{\cal Z} = - {\rm Tr} \int_0^\infty {ds\over s}
$$
\eqn\kapaaa{
\times \bigg\{ e^{-s\bigg[-\bigg(
\partial^{\mu}-ie\big(
e^{\overleftarrow{\partial}_\alpha \Theta^{\alpha \beta}
\overrightarrow{\partial}_\beta}A^\mu\big)\bigg) \bigg(\partial_\mu -ie \big(A_\mu
e^{\overleftarrow{\partial}_\alpha \Theta^{\alpha \beta}
\overrightarrow{\partial}_\beta}\big)\bigg) + M^2 -  {i \over 2} F_{\mu \nu}\bigg]}-
e^{-s} \bigg\}.
}
All exponential terms factorize and we can see that the problem reduces to the computation
for spin-${1\over 2}$ from the previous subsection plus the contribution of the factor 
${\rm tr} \exp \big( -{1 \over 2} s F^{\mu \nu}_{(j)}\big) = 2 \cosh(s x_j).$ 

Thus we get the amplitude $\widetilde{Z}$ by considering the ghost
contribution and it yields

\eqn\setaa{
\widetilde{\cal Z} = - \int_0^{\infty} {ds \over s} \prod_{j=1}^{2k+1} {1 \over 4
\pi}\bigg( {{1\over
2} x_j \over  \sinh ({s x_j \over 2})} \bigg)\bigg({1\over 1 + \Theta{x_j \over 4}}\bigg)
\bigg( -1 + \sum_{i=0}^{2k+1}2 \cosh (x_i) \bigg)
\exp(-sM^2).
}
After integrating out the $s$-variable we finally get

\eqn\setaunoa{
\widetilde{\cal Z}= - {1 \over (4 \pi)^{2k+1}} {1 \over M^2} \prod_{j=1}^{2k+1} {{1\over 2}
x_j \over
4 \pi \sinh ({1 \over 2}x_j )} \bigg({1\over 1 + \Theta{x_j \over 4}}\bigg)
\bigg(-1 + \sum_{i=0}^{2k+1}2 \cosh (x_i) \bigg).
}

Then, the total amplitude for Rarita-Schwinger fields is given by

\eqn\setadosa{
I_{3\over 2}(total) = -i {1 \over (2 \pi)^{2k+1}}  R(\varepsilon^{(i)}, p^{(j)})
 \prod_{j=1}^{2k+1} {{1\over 2} x_j \over
4 \pi \sinh ({1 \over 2}x_j )} \bigg({1\over 1 + \Theta{x_j \over 4}}\bigg)
\bigg( -1 + \sum_{i=0}^{2k+1}2 \cosh (x_i) \bigg).
}
Then, the total amplitude for Rarita-Schwinger fields is also modified by the same
$\Theta$-dependent factor justly as the spin-${1\over 2}$ fields of the previous section.
%%%%%%%%%%%%%%%%%%%%%%%%%%%%%%%%%%%%%%%%%%%%%%%%%%%%%%%%%%%%%%%%
%%%%%%%%%%%%%%%%%%%%%%%%%%%%%%%%%%%%%%%%%%%%%%%%%%%%%%%%%%%%%%%%  
\newsec{Noncommutative Mixed Anomalies}

%%%%%%%%%%%%%%%%%%%
%%%%%%%%%%%%%%%%%%%
\subsec{Mixed Anomaly for Spin ${1\over 2}$-Fields}

In this subsection we will compute mixed anomalies which include not only the
coupling of 
chiral fermions to gravity but also to nonabelian gauge fields. Noncommutative gauge
anomalies, for the case of Yang-Mills fields have been computed in a number of papers, see
for instance \refs{\gracia,\bonora,\bonoraone,\mink,\intri} for planar diagrams with gauge
group $U(N)$. 

We will consider a noncommutative spacetime of even dimension $D=2n$ and we compute
one-loop
amplitudes of $r$ external gluons and $n+1-r$ external gravitons. We will concentrate in
anomalous diagrams with $n+1-r=$even. Recent results concerning the computation of chiral
gauge anomalies in Yang-Mills theories in an even number of spacetime dimensions was performed
through the Wess-Zumino consistency condition in the reference \bonoraone. In the
present paper we apply the procedure of \bonora. For the case of non-planar diagrams there has
been some previous work in \refs{\ardalantwo,\esperanza,\nakajima}. The analysis can be
extended to other gauge
groups by introducing the Seiberg-Witten map as in references
\refs{\banerjee,\cpmartinsw,\rrcouplings,\brandt,\review}.

Before the evaluation of the relevant diagrams let us review briefly some ideas of
noncommutative Yang-Mills theory. Consider a gauge theory with a hermitian connection,
invariant under a symmetry Lie group ${\bf G}$, with gauge fields $A_\mu$ and gauge
transformations: $ \delta_{\lambda }{A}_{\mu} =\partial
_{\mu}{\lambda}+i\big[\lambda,{A}_{\mu}\big],$ with $\lambda=\lambda^iT_i$, and $T_i$ are
the generators of the Lie algebra ${\cal G}$ of the group ${\bf G}$, in the adjoint
representation. In the noncommutative Yang-Mills theory, the product of functions on the
spacetime manifold is promoted to the Moyal product. The above transformations are
generalized for the noncommutative theory as, $\delta_{\lambda }\widehat{A}_{\mu} =\partial
_{\mu}\widehat{\Lambda}+i\big[\widehat\Lambda,\widehat{A}_{\mu}\big]_*,$ where the
commutators are defined as $\big[A,B\big]_*\equiv A\ast B-B\ast A$. Due to
noncommutativity, a generic
commutator takes values in the universal enveloping algebra ${\cal U}({\cal G},{\bf R})$ of
the Lie algebra ${\cal G}$ in the representation ${\bf R}$ (for more details see, for instance
\refs{\wess}). In
particular,
$\big[\widehat\Lambda,\widehat{A}_{\mu}\big]_*$ take values in the
universal enveloping algebra ${\cal U}(su(N),{\bf ad})$ of the Lie algebra $su(N)$ (where,
for instance, 
${\bf G}=SU(N)$) in the
adjoint representation ${\bf ad}$. Therefore, $\widehat\Lambda$ and the gauge fields
$\widehat{A}_\mu$ will also take values in this algebra. Let us write for instance
$\widehat\Lambda=\widehat\Lambda^I T_I$ and $\widehat A=\widehat{A}^I T_I$, then,

\eqn\uns{
\big[\widehat\Lambda,\widehat{A}_{\mu}\big]_*=
\big\{\widehat{\Lambda}^{I},\widehat{A}_{\mu}^{J}\big\}_* \big[
T_{I},T_{J}\big] +\big[
\widehat\Lambda^{I},\widehat{A}_{\mu}^{J}\big]_*
\big\{ T_{I},T_{J}\big\},
}
where $\{A,B\}_*\equiv A\ast B+B\ast A$ is the noncommutative
anticommutator and the indices $I,J,K$ etc, run over the number of generators of the
enveloping algebra.
Thus all the products of the generators $T_I$ will be needed in order to close the algebra
${\cal U}({\cal G},{\bf ad})$.
Its structure can be obtained by successive computation of commutators and anticommutators
starting from the generators of ${\cal G}$, until it closes,

\eqn\due{
\big[ T_{I},T_{J}\big]=i{f_{IJ}}^KT_{K}, \ \ \ \ \big\{ T_{I},T_{J}\big\}
= {d_{IJ}}^KT_{K}.  
}
The field strength is defined as
$\widehat{F}_{\mu \nu} =\partial _{\mu}\widehat{A}_{\nu}-
\partial _{\nu}\widehat{A}_{\mu}-i
[\widehat{A}_{\mu},\widehat{A}_{\nu}]_*$,
hence
it takes also values in ${\cal U}({\cal G},{\rm ad})$. 

Thus axial anomalies in $2n$ dimensions can be obtained by computing
the amplitude associated to the one-loop diagram with $r$ external gluons and $n+1-r$ 
external gravitons. In the noncommutative case at each gluon vertex we have to insert

\eqn\gluonvtx{
-i \Gamma^{\mu} T^I_L \exp \big(-{i \over 2} \Theta^{\rho
\sigma} p_{1 \rho}p_{2 \sigma} \big) \delta(p_1 + p_2 + k),
}
where $T^I_L$ is the generator of the enveloping algebra ${\cal U}({\cal G},{\bf R})$ in
the
representation ${\bf R}$ furnished by the left-handed fermions. The group theory
factor associated with a given diagram is: ${\rm Tr} \big( T^{I_1}_L \cdot   T^{I_2}_L
\dots
T^{I_r}_L \big).$ After this factor is extracted, in each gluon vertex we have a factor
given by

\eqn\gluonvtx{
-i \Gamma^{\mu} \exp \big(-{i \over 2} \Theta^{\rho
\sigma} p_{1 \rho}p_{2 \sigma} \big) \delta(p_1 + p_2 + k).
} 
On the other hand in each graviton vertex we have to insert the factor

$$
-{i \over 4} \varepsilon^{\mu \nu} \Gamma_{\mu} \bigg({1 - \bar{\Gamma} \over 2} \bigg)
(p+p')_{\nu} 
\exp \bigg( -{i \over 2} \Theta^{\mu \nu} p_{\mu} p'_{\nu} \bigg).
$$

Dirac algebra of matrices $\Gamma$ can be carried out  and then trace does not
distinguish of the graviton and gluon vertices. Thus the kinematic factor
$R(\varepsilon^{(i)},p^{(j)})$ is exactly the same. After that, graviton vertex correspond
to a
massive complex scalar fields of charge ${1 \over 4}$ interacting with ``photons'' which
give rise
to a noncommutative effective theory of charged scalars coupled to external photons of the 
same type as that described in the previous section. After Dirac and group trace for the
gluon vertex \gluonvtx\ we have $ -i \exp \big(-{i \over 2} \Theta^{\rho
\sigma} p_{1 \rho}p_{2 \sigma} \big)$. This remaining noncommutative vertex corresponds to
a coupling of a scalar fields to the mentioned massive complex scalars. Thus the remaining
effective diagram consist of external  
scalar and photon fields coupled to complex scalar fields, with the usual propagators
$i/(p^2
-M^2)$, obeying noncommutative interactions. 

Similarly to the commutative case now we have to restrict trace formula to the symmetric
trace since noncommutativity respects the symmetry under permutations of external lines
\sheikh. Thus the factor ${\cal Z}'$ is given by

$$
{\cal Z}'(\Theta) = - {\rm STr} \bigg[ T^{I_1}_L T^{I_2}_L \dots  T^{I_r}_L \exp \bigg(-{i
\over 2} \Theta^{\rho \sigma} \sum_{\ell=1}^{r-1}p^{\ell}_{1 \rho}p^{\ell}_{2 \sigma} \bigg) 
\bigg]
$$
\eqn\cinc{
\times \bigg({\partial \over \partial
M^2}\bigg)^r
\int{ds \over s} \prod_{j=1}^{2k+1} \bigg[ {{1\over 2}
x_j \over
4 \pi \sinh ({sx_j\over 2})} {1\over \big(1 + \Theta {x_j \over 4}\big)}\bigg] \exp(-sM^2),
}
where the derivative stands, as in the commutative case, that the $-i \cdot  \exp
\big(-{i \over 2} \Theta^{\rho
\sigma} p_{1 \rho}p_{2 \sigma} \big)$ vertex has can be
obtained through a derivative with respect the mass square $M^2$, {\it i.e.}
${i \over p^2 - M^2}(-i) {i \over p^2 - M^2} = {\partial \over \partial M^2} \big[{i \over
p^2 - M^2}\big]$. Here ${\rm STr}$ is the symmetrized trace in the factor corresponding the
gauge amplitude is constructed by insering in each vertex a factor: $-i \Gamma^{\mu} \exp
\big(-{i \over 2} \Theta^{\rho \sigma} \ell^{(j)}_{\rho} \ell^{(j+1)}_{\sigma} \big)$. Then
the symmetrized trace is given by

$$
{\rm Tr} \big[ T^{I_1}_L T^{I_2}_L \dots  T^{I_r}_L \big] \bigg\{ \cos {\ell^{(1)} \Theta
\ell^{(2)} \over 2} \cdot \cos {\ell^{(3)} \Theta
\ell^{(4)} \over 2} \dots \cos {\ell^{(r-1)} \Theta
\ell^{(r)} \over 2} + {\rm all \ permutations} \bigg\},
$$
where $\ell^{(i)} \Theta \ell^{(i+1)} \equiv \Theta^{\rho \sigma} \ell^{(i)}_{\rho} 
\ell^{(i+1)}_{\sigma}.$

For instance for $r=4$ we have 

$$
{\rm Tr} \big[ T^{I_1}_L T^{I_2}_L T^{I_3}_L T^{I_4}_L\big] \bigg\{\cos {\ell^{(1)} \Theta
\ell^{(2)} \over 2} \cdot \cos {\ell^{(3)} \Theta
\ell^{(4)} \over 2} + \cos {\ell^{(1)} \Theta
\ell^{(3)} \over 2} \cdot \cos {\ell^{(2)} \Theta
\ell^{(4)} \over 2} 
$$
$$
+ \cos {\ell^{(1)} \Theta
\ell^{(4)} \over 2} \cdot \cos {\ell^{(2)} \Theta
\ell^{(3)} \over 2} \bigg\}.
$$

After $s$-integration we finally get that the total mixing anomaly is
given by

$$
I'_{1\over 2} = - {\rm STr} \bigg[ T^{I_1}_L T^{I_2}_L \dots  T^{I_r}_L \exp \big(-{i
\over 2} 
\Theta^{\rho \sigma} \sum_{\ell =1}^{r-1}p^{\ell}_{1 \rho}p^{\ell}_{2 \sigma} \big) \bigg]
$$
\eqn\sept{
\times {i \over (2 \pi)^{n\over 2}}  
R(\varepsilon^{(i)}, p^{(j)}) 
\prod_{j=1}^{n \over 2}{{1\over 2} x_j \over 4 \pi \sinh ({1 \over 2}x_j)}{1 \over \big(1
+ \Theta {x_j
\over 4}\big)}.
}
The interpretation of the gauge anomaly is justy as in case of chiral gauge anomaly in
Yang-Mills theory. In the case of $U(N)$ gauge group, as was described in Ref. \bonoraone, the
noncommutativity imposes more restrictive conditions for anomaly cancelation. Thus in order 
a noncommutative gauge theory be anomaly free this theory must be non-chiral. In four
dimensions noncommutative chiral gauge field theories with $U(N)$ group with adjoint matter is
anomaly free but is not loger true in higher dimensions. For instance in our present
case of $D=4k+2$ dimensions it has been showed \bonoraone\ that in for adjoint matter, chiral
anomaly is non-vanishing and it is precisely the $2N$ times the anomaly in the fundamental
representation.

%%%%%%%%%%%%%%%%%%%
%%%%%%%%%%%%%%%%%%%
\subsec{Mixed Anomaly for Spin-${3\over 2}$ Fields}

Similarly to the case of noncommutative mixed anomalies of gauge and gravitational
fields coupled to
complex chiral spin-${1\over 2}$ we we can compute the mixed anomalies for spin-${3
\over 2}$ case. Then we have

$$
\widetilde{\cal Z}'(\Theta) = - {\rm STr} \bigg[ \widetilde{T}^{I_1}_L
\widetilde{T}^{I_2}_L
\dots \widetilde{T}^{I_r}_L
\exp \big(-{i
\over 2} \Theta^{\rho \sigma} \sum_{\ell=1}^rp^{\ell}_{1 \rho}p^{\ell}_{2 \sigma} \big) 
\bigg]
$$
\eqn\cinc{
\times \bigg({\partial \over \partial
M^2}\bigg)^r
\int{ds \over s} \prod_{j=1}^{2k+1} \bigg[ {{1\over 2}
x_j \over
4 \pi \sinh ({sx_j\over 2})} {1\over \big(1 + \Theta {x_j \over 4}\big)}\bigg] \bigg(-1
+ \sum_{i=0}^{2k+1}  2 \cosh (x_i) \bigg) \exp(-sM^2).
}
After $s$-integration we finally get

$$
I'_{3\over 2} = - {\rm STr} \bigg[ \widetilde{T}^{I_1}_L \widetilde{T}^{I_2}_L \dots
\widetilde{T}^{I_r}_L \exp
\big(-{i
\over 2} 
\Theta^{\rho \sigma} \sum_{\ell =1}^rp^{\ell}_{1 \rho}p^{\ell}_{2 \sigma} \big) \bigg]
$$
\eqn\septagen{
\times {i \over (2 \pi)^{n \over 2}}  
R(\varepsilon^{(i)}, p^{(j)}) 
\prod_{j=1}^{n \over 2}{{1\over 2} x_j \over 4 \pi \sinh ({1 \over 2}x_j)}{1 \over \big(1
+ 
\Theta {x_j \over 4}\big)} \bigg( \sum_{i=1}^{n \over 2}  2 \cosh (x_i) \bigg),
}
where ${\rm STr}$ stands for the symmetrized trace which is defined as in the previous
subsection.

\vskip 1truecm
%%%%%%%%%%%%%%%%%%%%%%%%%%%%%%%%%%%%%%%%%%%%%%%%%%%%%%%%%%%%%%%%
%%%%%%%%%%%%%%%%%%%%%%%%%%%%%%%%%%%%%%%%%%%%%%%%%%%%%%%%%%%%%%%%
%%%%%%%%%%%%%%%%%%%%%%%%%%%%%%%%%%%%%%%%%%%%%%%%%%%%%%%%%%%%%%%%
\newsec{Final Remarks}

In this paper we have studied axial and chiral gravitational anomalies in the context of
noncommutative field theory. An interesting feature is that they are natural
higher-dimensional generalizations of the studies of axial and gauge anomalies in
noncommutative gauge theories. In order to compute the noncommutative effects we have used a
linearization of a noncommutative deformation of Einstein theory \moffatperturb, but in
principle, we could used any other noncommutative theory of gravity.  This noncommutative
deformation of linear gravity has been coupled to noncommutative chiral fermions and we have
assumed that gravity as well as matter are deformed with the same deformation parameter
$\Theta$. Thus, we focus on the interaction action of chiral fermions and the gravitational
field. We have provided the Feynman rules of this noncommutative theory, in particular
\ncfruleone\ was the necessary rule to determine the anomalies of planar diagrams. Anomalies
coming from non-planar diagrams were not considered in the present paper. The only
modification appears on the vertices of Feynman diagrams and we use them to compute a series
of processes involving gravitational anomalies.

After discussing the Feynman rules we have computed the noncommutative contribution to the
gravitational axial (ABJ) anomaly leading to the pion decay into two photons. This
noncommutative extension of the Delbourgo-Salam anomaly is obtained by using the dimensional
regularization method and we found that it gives precisely a noncommutative deformation of
signature $\widehat{\tau}(X)$ which is the spacetime analogous to the group signature worked
out in Ref. \nctopo.

As in the commutative case, noncommutative Delbourgo-Salam anomaly does not spoil
diffeomorphism or local Lorentz gauge invariance at the quantum level. However
noncommutativity might affect also these gauge symmetries as far as Lorentz transformations
and diffeomorphism symmetries are affected in noncommutative field theories.

In the two-dimensional case of the pure gravitational chiral anomaly, we have computed
diffeomorphism anomaly and we found that the noncommutativity does not
affect the
effective action $\Gamma(Q)$ and therefore the anomaly is the
same than in the usual commutative case obtained in \alwitt. This is also done in the general 
case of $D=4k+2$ dimensions. The anomaly was obtained by finding first a
noncommutative
residual interaction of a complex scalar field with an $U(1)$ gauge field. Here as usual in
the commutative
case, for each coupling
vertex we have translated the graviton and chiral fermion interaction to the problem to the 
problem of the vertex of a complex scalar field coupled to external non-dynamical
external photons. The effective action is
computed by using a two-dimensional noncommutative version Schwinger model. We find a
noncommutative
deformation of the effective action given by the expression \setados\ and \roof. The
computation of the anomaly for a loop of spin-${3\over 2}$ fields was performed and it was
obtained also a noncommutative correction given by the expression \setadosa.

Mixed anomalies also were computed in within this context and there is also a noncommutative
modification given by \sept\ and \septagen\ for spin-${1 \over 2}$ and spin-${3 \over 2}$
fields respectively.

There are several interesting points concerning the results of this work. One of them
concerns the application to the different ten-dimensional supergravities coming from string
theory. It would be interesting to compute the gauge and gravitational anomalies
due to anti-symmetric $p$-form fields in a noncommutative background and to
look for the conditions for the cancellation of these noncommutative anomalies in type I
and type II supergravities ten dimensions. Before of
solving these problem perhaps one first would address the problem of to give a sensible
theory of noncommutative extension of gauge theory for higher rank form potentials in higher
dimensions. 

Another interesting problem is the computation of gravitational anomalies due to non-planar
diagrams following \refs{\ardalantwo,\esperanza,\nakajima}. In the present paper we limited to
compute chiral gauge anomalies for $U(N)$ group. We would like to extend the computation to
other gauge groups by using the Seiberg-Witten map following some literature on this subject
\refs{\banerjee,\cpmartinsw,\rrcouplings,\brandt,\review}. We would like to apply the
Seiberg-Witten map for the gravitational sector as was discussed in
\refs{\chamone,\chamtwo,\nctopo,\ncsdg}.

One of the most interesting problems in to connect our results given by Eqs.  \setados, \roof
and \setadosa\ with the Atiyah-Singer index theorem for families of elliptic operators and to
give explicit formulas for these noncommutative anomalies in terms of the invariant
polynomials describing Pontrjagin and Chern characteristic classes. This is left for a future
communication. A description in terms of the Wess-Zumino consistency condition similarly to
\bonoraone\ it worth to provide for the case of gravity. In order to do that the
results of Ref. \calmet\ would be important. 

Finally, it would be very interesting also to pursue a suitable global approach, including
gravitational global anomalies \globo, and compare it with the results
given recently by Perrot \perrot\ in the computation of noncommutative gravitational anomalies
using different global tools.

\vskip 1truecm
%%%%%%%%%%%%%%%%%%%%%%%
\centerline{Acknowledgements}
This work is supported in part by a CONACyT M\'exico grant 33951E. C. S.-C. and  S.E. are
supported by a CONACyT graduate fellowship. H. G.-C. thanks O. Obreg\'on, R.
Rabad\'an and C. Ramirez by useful discussions. 

%%%%%%%%%%%%%%%%%%%%%%%
%%%%%%%%%%%%%%%%%%%%%%%
%%%%%%%%%%%%%%%%%%%%%%%
\vfill
\break

%%%%%%%%%%%%%%%%%%%%%%%%%%%%%%%%%%%%%%%%%%%%%%%%%%%%%%%%%%%%%%%%%%%%%%%%%%%
%%%%%%%%%%%%%%%%%%%%%%%%%%%%%%%%%%%%%%%%%%%%%%%%%%%%%%%%%%%%%%%%%%%%%%%%%%%

\listrefs

\end

The noncommutative roof-genus is given by

$$
\widehat{A}_{\Theta}(X)= 1 + \big(b_2 + {\Theta^2 \over 2}\big)(-{1 \over 2}) {\rm tr}
\big(R^2\big) + \big(b_4 + {\Theta^4 \over 24}\big)({1 \over 2}) {\rm tr}   
\big(R^4\big) + b_2\big(B_2 + {\Theta^2 \over 2}\big)({1 \over 4}) \big({\rm
tr}R^2\big)^2 
$$
$$
+ \big(b_6 + {\Theta^6 \over 720}\big)(-{1 \over 2}) {\rm tr}   
\big(R^6\big) + b_2\big(b_4 + b_4{\Theta^2 \over 2} + b_2 {\Theta^4 \over 24}\big)(-{1
\over 4}) {\rm tr}   
\big(R^2\big) \cdot {\rm tr}
\big(R^4\big) + \dots ,
$$
where $b_2 =-{1 \over 24}$, $b_4= ????$ and $b_6= ????$. It is worth to mention that the
roof-genus contributes with a $2k+2$